\documentclass[11pt]{article}
\pdfoutput=1
\usepackage[T1]{fontenc} 

\usepackage{verbatim}

\usepackage{bbm}
\usepackage{amsmath}
\usepackage{amssymb}
\usepackage{amsfonts}
\usepackage{mathrsfs}
\usepackage{slashed}

\usepackage[pdftex]{graphicx}
\usepackage[center,footnotesize,hang]{subfigure}
\usepackage{epsfig}
\usepackage{psfrag}
\usepackage{pstricks}
\usepackage{color}
\usepackage{color}
\usepackage{cancel}
\newcommand{\be}{\begin{equation}}
\newcommand{\ee}{\end{equation}}
\newcommand{\bea}{\begin{eqnarray}}
\newcommand{\eea}{\end{eqnarray}}

\usepackage{jheppub} 

\usepackage[T1]{fontenc} 

\leftline{\blue{ULB-TH/12-12\hspace{1.5cm} FTUAM-12-100 \hspace{1.5cm}IFT-UAM/CSIC-12-78}}
\bigskip

\title{\boldmath Muon conversion to electron in nuclei
in type-I seesaw models}

\author[a]{R. Alonso,}
\author[b]{M. Dhen,}
\author[a]{M. B. Gavela,}
\author[b]{T. Hambye}

\affiliation[a]{Departamento de F\'isica Te\'orica, Universidad Aut\'onoma de Madrid and\\ 
Instituto de F\'isica Te\'orica IFT-UAM/CSIC, Cantoblanco, 28049 Madrid, Spain}
\affiliation[b]{Service de Physique Th\'eorique, Universit\'e Libre de Bruxelles, \\
Bld du Triomphe, CP225, 1050 Brussels, Belgium}

\emailAdd{rodrigo.alonso@uam.es}
\emailAdd{mikadhen@ulb.ac.be}
\emailAdd{belen.gavela@uam.es}
\emailAdd{thambye@ulb.ac.be}

\abstract{ We compute the $\mu\rightarrow e$ conversion in the type-I seesaw model,  as a function of the right-handed neutrino mixings and masses. The results are compared with
 previous computations in the literature.
   We determine the definite predictions resulting for the ratios between the $\mu \rightarrow e$ conversion rate for a given nucleus and the rate of two other processes which also involve a $\mu-e$  flavour transition: $\mu \rightarrow e\gamma$ and $\mu\rightarrow eee$. For a quasi-degenerate mass spectrum of right-handed neutrino masses -which is the most natural  scenario leading to observable rates- those ratios depend only on the seesaw mass scale, offering a quite interesting testing ground. In the case of sterile neutrinos heavier than the electroweak scale, these ratios vanish typically for a mass scale of order a few TeV. 
 Furthermore, the analysis performed here is also valid down to very light masses. 
   It turns out that planned $\mu \rightarrow e$ conversion experiments would be sensitive to masses as low as $2$ MeV. 
   Taking into account other experimental constraints,  we show  that future $\mu \rightarrow e$ conversion experiments will be fully relevant to detect or constrain
sterile neutrino scenarios in the $2$ GeV$-1000$ TeV mass range. }
\keywords{Beyond Standard Model, Neutrino Physics, Rare Decays}

\begin{document}

\maketitle
\flushbottom

\section{Introduction}

The recent experimental evidence for neutrino masses has shown that lepton flavour is violated in the neutrino sector.
This inevitably leads, at the tree or one-loop level, to rare processes violating charged lepton flavour, such as $l\rightarrow l'\gamma$, $l\rightarrow 3 l'$ or $\mu$~to~$e$ conversion in atomic nuclei. Current experimental bounds on the various rates are expected to be improved in the near future by a long series of new experiments. In particular, $\mu\rightarrow e$ conversion processes~\cite{Weinberg:1959zz} will become especially competitive, as the sensitivities for various nuclei are expected to be improved by several orders of magnitude, 

\begin{eqnarray}
R_{\mu\rightarrow e}^{Ti}&\lesssim10^{-18}& \quad \mbox{\cite{Hungerford:2009zz,Cui:2009zz}}\, ,
\label{Tiexpected}\\
R_{\mu\rightarrow e}^{Al}&\lesssim 10^{-16}&\quad \mbox{\cite{Hungerford:2009zz,Cui:2009zz,Carey:2008zz, Kutschke:2011ux,Kurup:2011zza}}\,,
\label{Alexpected}
\end{eqnarray}
as compared to the present sensitivities 
\begin{eqnarray}
R_{\mu\rightarrow e}^{Ti}&<& 4.3 \times 10^{-12}\quad \mbox{\cite{Dohmen:1993mp}} \, ,
\label{Tipresent}\\
R_{\mu\rightarrow e}^{Au}&<&7   \times 10^{-13}\quad \mbox{\cite{Bertl:2006up}}\, ,
\\
R_{\mu\rightarrow e}^{Pb}&<& 4.6 \times 10^{-11}\quad \mbox{\cite{Honecker:1996zf}}\,.
\end{eqnarray} 


The rates predicted by neutrino mass models are in general expected to be  far below these sensitivities. However this is not necessarily the case. The rates are highly model dependent. Classes of models, such as those based  on approximate conservation of lepton number (predicting 2 or more quasi-degenerate heavy fermions), can naturally give large flavour-changing rates and the experimental sensitivities can be saturated without fine-tuning.

The $\mu\rightarrow e$ conversion rate $R_{\mu\rightarrow e}$ has been calculated in the literature for the various possible types of seesaw, with right-handed neutrinos~\cite{Riazuddin:1981hz,Chang:1994hz,Ioannisian:1999cw,Pilaftsis:2005rv,Deppisch:2005zm,Ilakovac:2009jf,Deppisch:2010fr}, scalar triplet(s)~\cite{Raidal:1997hq,Ma:2000xh,Dinh:2012bp} and fermion triplets~\cite{Abada:2008ea}. 
For the right-handed neutrino case ("type-I" seesaw), a comparison of the various calculations shows that there is no agreement on what is actually the result. This issue is also relevant for references using calculations in previous articles~\cite{Dinh:2012bp,AristizabalSierra:2012yy}.  Some of the discrepancies have a very significant impact on the predictions. Given both these experimental and theoretical situations, in this paper we recalculate the $\mu\rightarrow e$ conversion rate for the  type-I seesaw model. We compare our results with previous calculations and determine the corresponding phenomenology it leads to.  

In the phenomenological analysis, we begin by comparing the conversion rate with the $\mu \rightarrow e \gamma$ and $\mu \rightarrow eee$ rates; the latter process is very close to $\mu \to e $ conversion in the sense that both transitions involve the same (local and non-local) form factors and consequently they will be shown to share common features  in the asymptotic regimes; the former instead depends only on the dipole form factor for real photons, which also gives a non-local contribution to $\mu \to e$ conversion  and $\mu \rightarrow eee$ decay.

For a quasi-degenerate spectrum of right-handed neutrinos, a case which can naturally allow large rates, it turns out that the ratio of two $\mu$-$e$ processes ($\mu\rightarrow e$  in Titanium, $\mu\rightarrow e$  in Aluminium, $\mu \rightarrow e \gamma$, $\mu \rightarrow eee$, ...) depends only on the right-handed neutrino mass scale~\cite{Chu:2011jg}, offering a quite interesting testing ground. 
As a function of that scale, we determine the $R_{\mu \rightarrow e}/Br(\mu\rightarrow e \gamma)$, $R_{\mu \rightarrow e}/Br(\mu\rightarrow eee)$ ratios for Aluminium, Titanium, Lead and Gold nuclei, and also the $R_{\mu \rightarrow e}^{Ti}/R_{\mu \rightarrow e}^{Al}$ ratio.
The observation of any two or more $\mu$-$e$ transitions, in agreement with the predicted ratios, would provide a strong evidence for such type-I seesaw scenarios.

 In this paper we also determine the range of right-handed neutrino masses and mixings the future $\mu \rightarrow e$ conversion experiments could reach. 
Although models with  light seesaw scales may be unattractive theoretically as they tend to require large fine-tunings, it is worth to explore the full range of scales open-mindedly, as the results will apply to any singlet fermion in nature.  It will be shown that the reach of planned $\mu\rightarrow e$ conversion experiments taken by themselves extends  from $1000$ TeV down to the MeV scale.
Furthermore, for sterile species lighter than the electroweak scale, experimental bounds on the non-unitarity of the leptonic mixing matrix in the electron and muon sectors~\cite{Antusch:2006vwa,Antusch:2008tz} become relevant and will be taken into account. 
For right-handed neutrinos lighter  than $\simeq 2$~GeV, the constraints on sterile-electron/muon coupling from  $K$ and $D$ decays,  as well as other constraints, 
 superseed the sensitivity  of $\mu\rightarrow e$ conversion processes.
   It will be shown that the planned $\mu\rightarrow e$ conversion experiments may  become the main actor in detecting or setting constraints on  sterile neutrino scenarios  
 for mass scales from $1000$ TeV down to $2$ GeV.

 Sect.~\ref{lagrangian} describes the Lagrangian and couplings. Sect.~\ref{generalities}  presents our calculation and deals with the general behaviour of the rates for the different processes considered, for any value of the right-handed neutrino masses; it also considers the particular cases of quasi-degenerate and hierarchical spectra. In Sect.~\ref{largemN} the regime of right-handed neutrino masses larger than the electroweak scale is discussed, both analytically and numerically, and the corresponding phenomenological analysis is performed; it also includes a comparison with previous computations in the literature. Sect.~\ref{lowmN} focuses on the analogous analysis for heavy neutrino masses lighter than the electroweak scale, and in Sect.~\ref{conclusions} we conclude. Exact analytical formulae and other items are included in the Appendix.

\section{Type-I seesaw model Lagrangian in mass eigenstate basis}
\label{lagrangian}
In the flavour basis, the Lagrangian of the type-I seesaw model with extra right-handed neutrinos $N_R$ reads
\begin{equation}
\mathcal{L}=\mathcal{L}_{SM}+i\overline{N_R}\slashed \partial N_R-\left[\overline{N_R}Y_N\tilde{\phi}^\dagger \ell_L +\frac{1}{2}\overline{N_R}M {N_R}^c+h.c.\right] \,,
\end{equation}
 where the flavour indices have been left implicit. 
Here $Y_N$ denotes the neutrino Yukawa coupling to the Standard Model (SM) scalar boson (so-called "Higgs boson" for short), $\ell_L$ denotes the left-handed lepton doublet, $\phi$ is the higgs doublet and $\tilde\phi=i\tau_2\phi^*$  with
$\tau_2$ the second Pauli matrix,
\begin{equation}
\ell_L=\left(\begin{array}{c}
\nu_L\\
\mathit{l_L}
\end{array}
\right)\,,
\qquad \qquad \phi=\left(\begin{array}{c}
\phi^-\\
\frac{1}{\sqrt{2}}\left(v+h+i\phi^3\right)
\end{array}\right)\,,
\end{equation}
and $v= 246$~GeV. The charged lepton Yukawa coupling is assumed diagonal without loss of generality. We define the unitary $(3+k)\times (3+k)$ mixing matrix $U$ through
\begin{equation}
\left(
\begin{array}{c}
\nu_L\\
{N_R}^{c}\\
\end{array}
\right)= U \, P_L\, n \equiv U \, P_L\,
 \left(\begin{array}{c} \nu_{1}\\ \nu_{2}\\ \nu_3\\N_1\\ \vdots \\ N_k\\ \end{array}\right)\,,
\label{Udef}
\end{equation}
where the vector $n$ encodes all neutrino eigenstates in the mass basis, with $N_i$, $i=1$ to $k$ denoting the $k$  extra physical heavy neutrinos, and  
 $P_L$ is the left-handed projector $P_L\equiv(1-\gamma_5)/2$. These mass eigenstates are Majorana fermions, $n=n^c$, that is $\nu_i= \nu_i^c$ and $N_i= N_i^c$. 
 In the mass eigenstate basis, the various gauge boson and scalar interactions read (for an arbitrary gauge, and  with $\alpha,\beta$ denoting the three flavour indices, and $i,j$ denoting the $3+k$ mass eigenstates)
\begin{eqnarray}
\mathcal{L}^{W^{\pm}} &=&   \frac{g_W}{\sqrt{2}} W^-_\mu \ \overline{l}_\alpha \gamma^\mu U_{\alpha i} P_L n_i + h.c. \,,\\
\mathcal{L}^{Z} &=&   \frac{g_W}{2 c_W} Z_\mu \ \overline{n}_i \gamma^\mu C_{ ij} P_L n_j = \frac{g_W}{4 c_W} Z_\mu \ \overline{n}_i \gamma^\mu \left[C_{ ij} P_L - C^*_{ ij} P_R\right] n_j \,,\\
\mathcal{L}^{\phi^{\pm}} &=& - \frac{g_W}{\sqrt{2}M_W} \phi^-  \ \overline{l}_\alpha   U_{\alpha i} \left(m_{l_\alpha} P_L - m_{n_i} P_R \right) n_i + h.c.\,, \\
\mathcal{L}^{\phi^3} &=&   -  \frac{i g_W}{2 M_W} \phi^3 \ \overline{n}_i   C_{ ij} \left(   m_{n_i} P_L -m_{n_j}   P_R   \right) n_j \,, \\
&=& -  \frac{i g_W}{4 M_W} \phi^3 \ \overline{n}_i   \left[ C_{ ij} \left(  m_{n_i} P_L -m_{n_j}   P_R \right)-C^*_{ ij} \left(  m_{n_i} P_R -m_{n_j}   P_L \right)\right]  n_j \,, \\ 
\mathcal{L}^{h} &=&      -\frac{  g_W}{2 M_W} h \ \overline{n}_i    C_{ ij} \left(  m_{n_i} P_L +m_{n_j}   P_R \right)  n_j \,,\\
&=&-\frac{  g_W}{4 M_W} h \ \overline{n}_i   \left[ C_{ ij} \left(  m_{n_i} P_L +m_{n_j}   P_R \right)+C^*_{ ij} \left(  m_{n_i} P_R +m_{n_j}   P_L \right)\right] n_j \,,
\end{eqnarray}
where $g_W$ is the weak isospin coupling constant, $c_W$ is the cosine of the weak mixing angle, and $C$ and $m_n$ are  $(3+k)\times (3+k)$ matrices defined as:
\begin{equation}
C_{ij}\equiv\sum_{\alpha=1}^3 U^\dagger_{i \alpha }U_{\alpha j }\,,\qquad \qquad m_n=\mbox{Diag}\left(m_{n_i}\right)=\mbox{Diag}(m_{\nu_1},m_{\nu_2},m_{\nu_3},m_{N_1},...,m_{N_k})\,.
\end{equation}
The first three entries in $m_n$ denote the light neutrino masses, while the last $k$  ones correspond to the heavy ones.
Although the general results to be shown below will not assume any expansion in powers of  $Y_N v/m_N$, 
 for all plots we will neglect higher order contributions in this expansion. For this case it is worth to decompose the $U$ matrix in blocks as
\begin{equation}
U=
\begin{pmatrix}
		U_{\nu\nu}& U_{\nu N} \\
		U_{N\nu}  & U_{NN}
\end{pmatrix}\,,
\end{equation}
which allows to recover the usual expressions valid  up to ${\cal O}[(Y_Nv,m_l)/m_N]^2$ corrections:
\begin{equation}
U_{\nu\nu}=(1-\frac{\epsilon}{2})U_{PMNS}\,\,\,,\quad U_{\nu N}=Y_N^\dagger (M)^{-1}\frac{v}{\sqrt{2}}\,\,\,,\quad U_{N\nu}=-M^{-1}Y_N \frac{v}{\sqrt{2}} U_{\nu\nu}\,\,\,,\quad U_{NN}=(1-\frac{\epsilon'}{2})\,,
\label{approxUs}
\end{equation}
where $\epsilon\equiv\frac{v^2}{2}Y_N^\dagger M^{-2}Y_N$, $\epsilon'\equiv\frac{v^2}{2} M^{-1}Y_N Y_N^\dagger M^{-1}$, and $U_{PMNS}$ denotes the customary lowest-order $3\times 3$ leptonic mixing matrix appearing in charged currents. With these conventions, the neutrino mass matrix is given by:
\begin{equation}
{\cal M}_\nu=-\frac{v^2}{2}Y_N^T\frac{1}{M}Y_N\,.
\end{equation}

\section{$\mu$ to $e$ conversion rates\label{muegammageneral}}
\label{generalities}

\subsection{Calculation of the rates}

In the type-I seesaw framework, violation of charged lepton number arises at the one loop level. $\mu$ to $e$ conversion 
is induced by a series of gauge boson mediated diagrams given in Fig.~\ref{diagrams}.
The various contributions
to the process can be divided in those in which the momentum is transferred by the
photon, by the Z boson or via two W bosons. The first two proceed via penguin
diagrams, whereas the latter processes  corresponds to a box diagram. Alike to the quark case, the internal fermions in the loop must have non-degenerate masses and non trivial mixings, in order to avoid a GIM cancellation. 
\begin{figure}[b!]
\centering
\subfigure[Photon Penguin Diagram]{\includegraphics[width=0.3\textwidth]{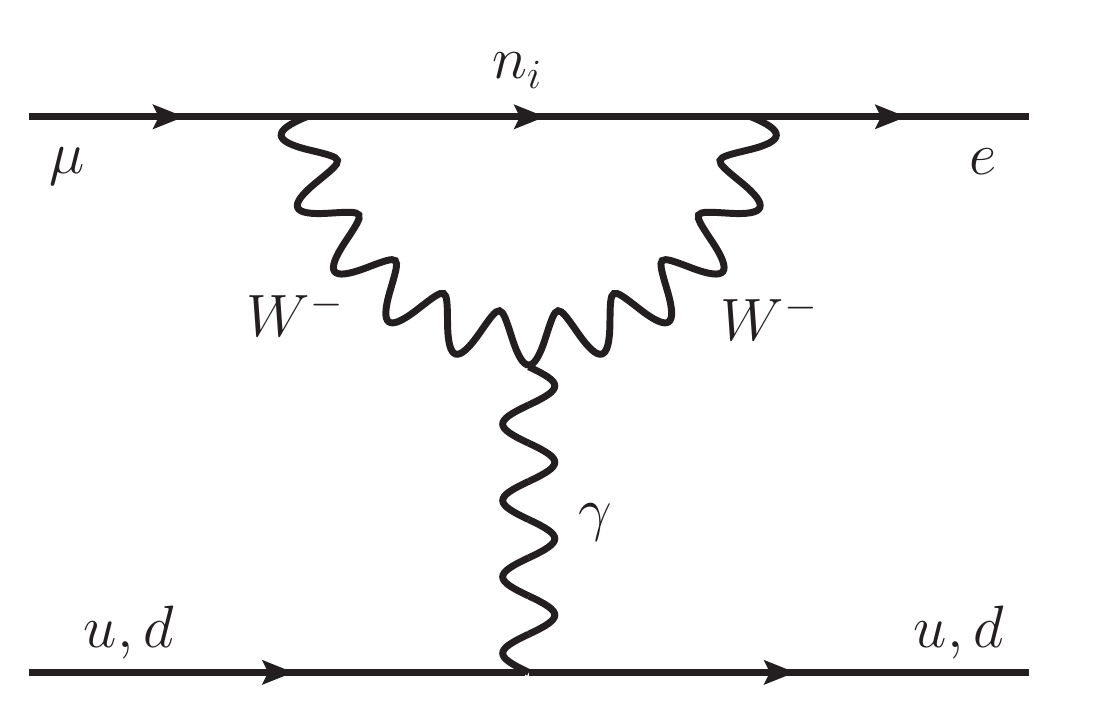}}
\subfigure[Z Penguin Diagram]{\includegraphics[width=0.3\textwidth]{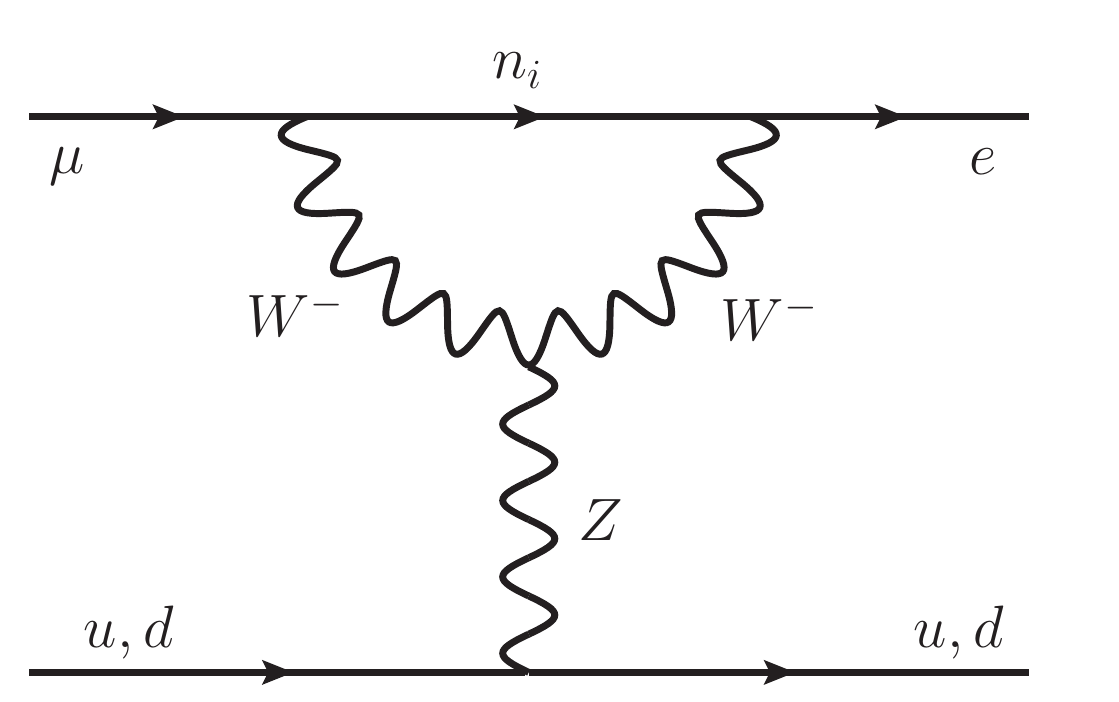}}
\subfigure[Z Penguin Diagram]{\includegraphics[width=0.3\textwidth]{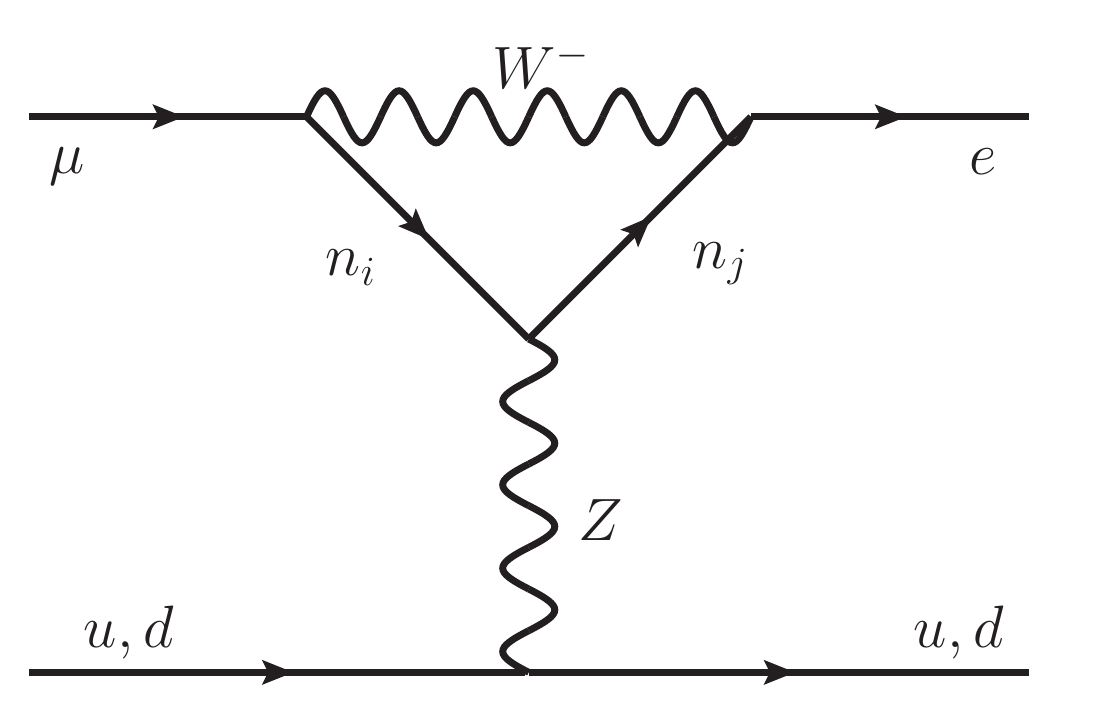}}
\subfigure[Box Diagram]{\includegraphics[width=0.3\textwidth]{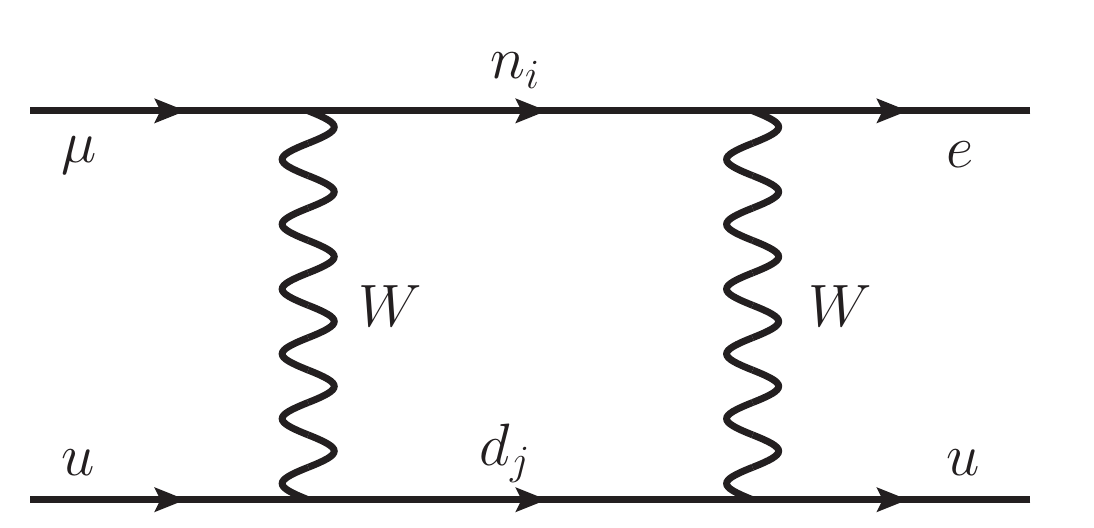}}
\subfigure[Box Diagram]{\includegraphics[width=0.3\textwidth]{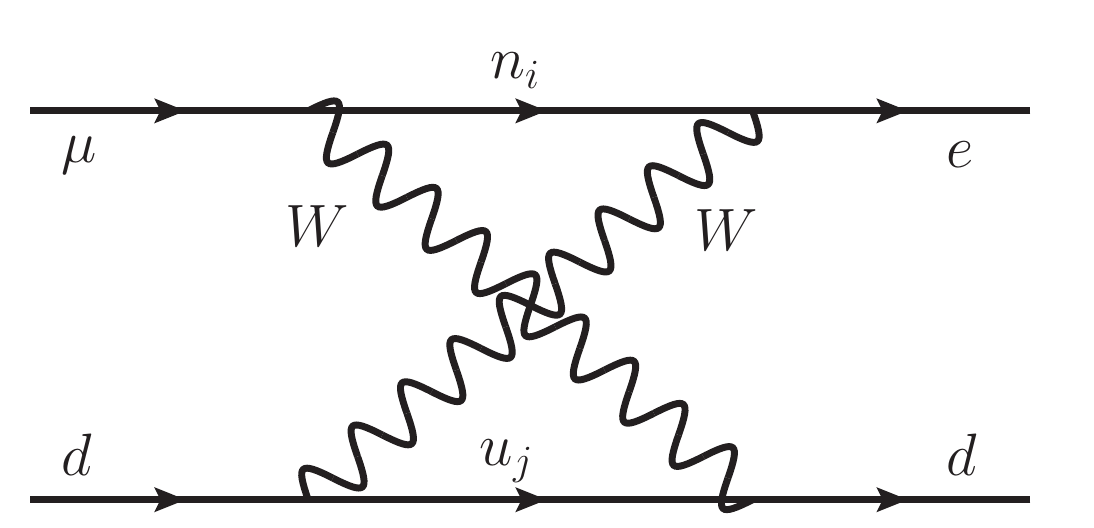}}
\label{diagrams}
\caption{The five classes of diagrams contributing to $\mu$ to $e$ conversion in the type-I seesaw model.}
\label{diagrams}
\end{figure}

For a rigorous calculation of the rate it is necessary to separate the local contributions from the "extended" ones. This stems from the fact that extended contributions, unlike local ones, are sensitive to atomic electric field effects. The W and Z mediated diagrams are obviously all local. The $\gamma$ mediated diagrams contribute to both classes of transitions, extended and local. The $\mu\rightarrow e \gamma$  matrix element can be written as
\begin{equation}
  i  \mathcal{M}=  \frac{ ieg_W^2}{2 (4\pi)^2 M^2_W} \epsilon^\mu_\lambda(q) \overline{u}_e(p')\Big[ F^{\mu e }_\gamma (q^2\gamma_\mu - {\not}q q_\mu) P_L   - i \sigma_{\mu\nu} q^\nu G^{\mu e}_\gamma (m_eP_L+m_\mu P_R) \Big] u_\mu(p)\,,
    \label{muegamma}
\end{equation}
where $q$ denotes the photon momentum, $q=p-p'$. The second term in this equation -mediated by the photon-lepton "dipole" $G_\gamma^{\mu e}$ coupling- is the only one contributing  for an on-shell photon and  is non local, whereas the "monopole" term $F_\gamma^{\mu e}$ is "local" (i.e.~it only accounts for off-shell photon exchange and it involves 2 powers of the photon momentum in the numerator which compensate the long range $1/q^2$ propagator of the photon between the lepton and nuclei lines~\cite{Czarnecki:1998iz}).
One can therefore divide the effective Lagrangian relevant for $\mu\rightarrow e$ conversion in two parts~\footnote{An interaction of the lepton current with
the axial quark current is also generated but negligible compared with the dominant coherent $\mu$ to $e$ conversion, $\mathcal{O}(A)$ larger.}:
\begin{equation}
\mathcal{L}_{eff}=\frac{eg_W^2m_\mu}{4(4\pi)^2M_W^2} G^{\mu e}_\gamma \bar e \,\sigma_{\lambda \rho}  \mu_R F^{\lambda\rho}
+\frac{g_W^4}{2(4\pi)^2M_W^2}\sum_{q=u,d}  \tilde{F}_q^{\mu e}\,
\bar e \,\gamma_\rho \mu_L\, \bar q \gamma^\rho  q+h.c.\,, 
\label{efflag}
\end{equation}
 where $F^{\lambda\rho}=\partial^\lambda A^\rho-\partial^\rho A^\lambda$ denotes the electromagnetic field strength, 
  and $\tilde{F}_q^{\mu e}$ contains the contribution of the monopole $F_\gamma^{\mu e}$ term as well as that from the weak gauge-boson exchange diagrams. The coefficients $\tilde F_q^{\mu e}$ and $G^{\mu e}_\gamma$
encode all the dependence on the internal fermion masses
and mixing angles.  In Eq.~(\ref{efflag}), and all through the rest of the paper, the electron mass has been neglected.

The effects of the nuclear form factors and of  averaging over the atomic electric field can be taken into account in the way described in Ref.~\cite{Kitano:2002mt}. 
The nuclear information is encoded by  $D$, $V^{(p)}$ and $V^{(n)}$ form factors whose values are shown in Table~1, taken from Ref.~\cite{Kitano:2002mt}. The final expression for the ratio of the $\mu\rightarrow e$ conversion over the capture rate $\Gamma_{capt}$  reads 
\begin{equation}
R_{\mu \rightarrow e}=\frac{2G_F^2\alpha_{w}^2m_\mu^5}{(4\pi)^2 \Gamma_{capt}
}\left|4V^{(p)}\left(2 \tilde{F}_{u}^{\mu e}+\tilde{F}_{d}^{\mu e}\right)+4V^{(n)}\left(\tilde{F}_u^{\mu e}+2\tilde{F}_{d}^{\mu e}\right)+ s^2_w G^{\mu e}_{\gamma} D/(2e)  \right|^2
\label{BRmueexact}
\end{equation}
with, as usual, $\alpha_W=g_W^2/(4\pi)$, $\alpha=e^2/(4\pi)$,  and $G_F$ the Fermi constant. The expression for the dipole term $G^{\mu e}_\gamma$ is given in the Appendix,  Eq.~(\ref{Ggammamue}). The form factors $\tilde{F}_{q}^{\mu e}$ are in turn given by  
\begin{equation}
\tilde F_q^{\mu e}=Q_q s_W^2 F^{\mu e}_\gamma+F^{\mu e}_Z\left(\frac{{\mathcal I}^3_q}{2}-Q_qs_W^2\right)+\frac{1}{4}F^{\mu eqq}_{box}\,,
\label{ay}
\end{equation}
where $q=u,d$, $Q_q$ is the quark electric charge ($Q_u=2/3, Q_d=-1/3$), $\mathcal{I}^3_q$ is the weak isospin ($\mathcal{I}^3_u=1/2\,,\, \mathcal{I}^3_d=-1/2$) and $s_W$ is the sinus of the weak mixing angle.   $F^{\mu e}_\gamma$, $F^{\mu e}_Z$ and $F^{\mu eqq}_{box}$ denote form factors corresponding to photon-penguin diagrams, $Z$-penguin diagrams and box diagrams in Fig.~1.

In the following we will use Eq.~(\ref{BRmueexact}) for all numerical results.
For low atomic number, $\alpha Z\ll1$, and for comparison with other calculations, the results can be nevertheless simplified. In this case  all the interactions can be considered point-like~\footnote{The virtuality of the photon is set by the transferred momentum $q^2=-m_\mu^2$: the interaction can be considered point-like if this scale is small compared to the 
Bohr radius $\propto ({Z\alpha})^{-1}$.} ($D/8e\sim V^{(p)} $) and the proton and neutron densities are approximately equal ($V^{(p)}/Z\sim V^{(n)}/(A-Z) $), which gives
\begin{equation}
R_{\mu \rightarrow e}=\frac{G_F^2 \alpha_W^2\alpha^3m_\mu^5}{8\pi^4\Gamma_{capt}}\frac{Z_{eff}^4}{Z}F_p^2  \left|
 Z\left(2 F_{u}^{\mu e}+F_{d}^{\mu e}\right)+(A-Z)\left(F_{u}^{\mu e}+2F_{d}^{\mu e}\right)\right|^2 \,.
 \label{BRmueapproxlight}
\end{equation}
In this equation, $A$ stands for the mass number, $Z_{eff}$ denotes the effective atomic number,  $F_p$ is a nuclear
form factor related to $V^{(p)}$ through $V^{(p)}/\sqrt{Z}\simeq Z_{eff}^2 F_p\alpha^{3/2}/4\pi  $, while $F_{(q)}^{\mu e}\equiv\tilde{F}_{(q)}^{\mu e}+Q_qs_W^2G^{\mu e}_\gamma$. The values of $Z_{eff}$, $F_p$ and $\Gamma_{capt}$ that will be used in the phenomenological analysis below are given in Table.~1, taken from Refs.~\cite{Kitano:2002mt} and~\cite{Suzuki:1987jf}.

\begin{table}[t]
\centering
\begin{tabular}{c | c | c | c | c |c |c }
\hline\hline
Nucleus $^A_Z \mbox{N}$          &$V^{(p)}$& $V^{(n)}$& $D$      & $Z_{eff}$   & $|F_p(-m^2_\mu)|$ &$ \Gamma_{capt} $ ($10^6s^{-1}$)\\
\hline
$_{13}^{27}\mbox{Al}$ & 0.0161   & 0.0173    & 0.0362 & $11.5$        &$0.64$                       &  0.7054\\
$_{22}^{48}\mbox{Ti}$ & 0.0396   & 0.0468    & 0.0864 & $17.6$         &$0.54$                      & 2.59\\
$^{197}_{79} \mbox{Au}        $  & 0.0974   & 0.146      & 0.189    & $33.5$         &$0.16$                     & 13.07\\
$^{208}_{82} \mbox{Pb }       $  & 0.0834   & 0.128      &  0.161   & $34.0$         &$0.15$                     & 13.45\\
\hline
\end{tabular} 
\label{nuclfactors}
\caption{Nuclear form factors and capture rates.}
\end{table}

Our complete analytical results for the amplitude of the various diagrams, together with their sum, are given in Eqs.~(\ref{Fmue}-\ref{Fmueee}) of the Appendix. 
The only approximations used in our calculation are to neglect:  i) 
the electron mass compared to the muon mass; ii)  higher orders in the external momentum over the W mass, as usual; iii)  the value of the three light neutrino masses compared to the extra $k$ heavier ones~\footnote{The latter approximation is not made in the first equality of Eqs.~(\ref{Fmue})-(\ref{Fmueee}),  only in the second equality of these equations. One could therefore use those results for any value of the right-handed neutrino masses.}.  The first 2 approximations are  accurate at a level better than ${\cal O}(10^{-4})$, that is to say better  that what can be expected from  higher loop contributions, and better than the uncertainties on the nuclear form factors. The third one  becomes excellent as soon as the right-handed neutrino masses are a few orders of magnitude above the  light neutrino masses.

Finally, when comparing the strength of $\mu\rightarrow e$ conversion processes with the branching ratio for $\mu\rightarrow e \gamma$ we will use for the latter the well-known result~\cite{Minkowski:1977sc}-\cite {Cheng:1980tp} 
\begin{equation}
 Br(\mu \to e \gamma)=\frac{\alpha^3_W s^2_W}{256 \pi^2}\frac{m^4_\mu}{M^4_W}\frac{m_\mu}{\Gamma_\mu}\big| G^{\mu e}_\gamma \big|^2 \,,  
    \label{muegexactrate}
\end{equation}
 where $\Gamma_\mu\approx 2.996\ 10^{-19}$ GeV denotes the total decay rate of the muon.  When comparing instead  with the strength of $\mu\rightarrow eee$ decay,  the expression for  the branching ratio $Br(\mu \to eee)$ will be taken from Ref.~\cite{Ilakovac:1994kj},
\begin{align}
Br(\mu   \rightarrow eee)=& \frac{\alpha^4_w }{24576\pi^3}\frac{m^4_\mu}{M^4_W}\frac{m_\mu}{\Gamma_\mu} \nonumber \\
		&\times \Bigg\{ 2 \left|\frac{1}{2}F^{\mu eee}_{Box}+F^{\mu e}_Z-2s^2_w(F^{\mu e}_Z-F^{\mu e}_\gamma)\right|^2+4 s^4_w \left|F^{\mu e}_Z-F^{\mu e}_\gamma\right|^2 \nonumber \\
		&+ 16 s^2_w Re\left[	(F^{\mu e}_Z +\frac{1}{2}F^{\mu eee}_{Box})	G^{\mu e*}_\gamma 			\right]		- 48 s^4_w Re\left[	(F^{\mu e}_Z-F^{\mu e}_\gamma)	G^{\mu e*}_\gamma 			\right]	\nonumber 		\\
		&+32 s^4_w |G^{\mu e}_\gamma|^2\left[		\ln \frac{m^2_\mu}{m^2_{e}} -\frac{11}{4}		\right]		\Bigg\}\,,  \label{mueee}
\end{align}
 see the Appendix for the values of the various form factors.
The comparison of this equation with Eqs.~(\ref{BRmueexact}) and (\ref{ay}) illustrates that $Br(\mu   \rightarrow eee)$ and $\mu \to e$ conversion are sensitive to different combinations of the same form factors.

\subsection{Phenomenological framework}

The $\mu-e$ transition rates  expected in the type-I seesaw  framework are clearly highly model dependent. 
 For right-handed neutrino masses  above the eV scale -as those considered in this work- for which the seesaw approximation  $U_{\nu N}
 \sim Y_N^\dagger v/M$  in Eq.~(\ref{approxUs}) holds, they scale in particular as the inverse of the right-handed neutrino masses at the fourth power, and contain four Yukawa couplings in the numerator, bringing each one a flavour dependence. 
Nevertheless it turns out that the models which can naturally give measurable rates are models which involve two  or more quasi-degenerate right-handed neutrinos and for these models one can make remarkably clear predictions.
The quasi-degenerate case is particularly natural, as it takes place for instance in scenarios in which lepton number (L) is approximately conserved~\cite{Wyler:1982dd,Mohapatra:1986bd,Branco:1988ex,GonzalezGarcia:1988rw,Barbieri:2003qd,Raidal:2004vt,Kersten:2007vk,Shaposhnikov:2006nn,Abada:2007ux,Asaka:2008,Ilakovac:1994kj,Gavela:2009cd}: the  degeneracy is protected by the  symmetry. This assumption allows a natural decoupling between large  Yukawa couplings (inducing L-preserving large rates) and small Yukawa couplings (guaranteeing small neutrino masses) 
and results in viable scenarios, even for low seesaw scales. 

 The remarkable predictions resulting  for the quasi-degenerate case hold for the ratios of two rates where a same flavour transition occurs, for example $R_{\mu\rightarrow e}/Br(\mu\to e \gamma)$ or  $R_{\mu\rightarrow e}/Br(\mu\to e e e)$. 
The point is simply that if the right-handed neutrinos are quasi degenerate, only one right-handed neutrino mass scale is relevant in the rates and the dependence on the elements of the mixing matrix (which contain the Yukawa dependence) factorizes from the mass dependence. For instance, for  $m_{N_1}\simeq m_{N_2}\simeq...\equiv m_N$, the rates in Eq.~(\ref{BRmueapproxlight}) for $\mu\to e$ conversion in light nuclei and Eq.~(\ref{muegexactrate}) for $\mu\rightarrow e\gamma$  [the generalization to heavy nuclei is straightforward from Eqs.~(\ref{BRmueexact}) and (\ref{BRmueapproxlight})] take the factorized form 
\begin{eqnarray} 
    R_{\mu\rightarrow e}&\simeq&  \frac{G_F^2 \alpha_W^2\alpha^3m_\mu^5}{8\pi^4\Gamma_{capt}}\frac{Z_{eff}^4}{Z}F_p^2   \nonumber    \label{RmuegammaPheno} \\
    && \times \Big[\left(A+Z\right)F_u(x_{N})+\left(2A-Z\right)F_d(x_{N})\Big]^2 \ \ \Big|\sum_{i}^{k} U_{eN_i}U^*_{\mu N_i}\Big|^2  \,. \label{muerate}\\
     Br(\mu \to e \gamma)&\simeq& \frac{\alpha^3_W s^2_W}{256 \pi^2}\frac{m^4_\mu}{M^4_W}\frac{m_\mu}{\Gamma_\mu}  \  G^2_{ \gamma}(x_{N}) \ \ \Big|\sum_{i}^{k} U_{eN_i}U^*_{\mu N_i}\Big|^2  \,,  \label{RmuePheno}
\end{eqnarray}
  where $G_\gamma(x_N)$, $F_u(x_N)$ and $F_d(x_N)$ are  functions of $x_N\equiv m^2_N/M^2_W$ given in Appendix, Eqs.~(\ref{Ggamma}), (\ref{functionfu}) and (\ref{functionfd}).  
In consequence, the Yukawa coupling (flavour) dependence drops in the ratio of both rates, leaving only a dependence on the mass $m_N$ of the heavy neutrinos~\cite{Chu:2011jg}:
\begin{equation}
    R^{\mu-e}_{\mu\to e\gamma}(x_N) \equiv \frac{R_{\mu\rightarrow e}}{Br(\mu\to e \gamma)} = 16\alpha^2_w\alpha F^2_p \frac{Z^4_{eff}}{Z}\frac{\Gamma_\mu}{\Gamma_{capt}}\left[\frac{ \left(A+Z\right)F_u(x_{N})+\left(2A-Z\right)F_d(x_{N}) }{G_\gamma(x_N)}\right]^2\,.
    \label{ratio1}
\end{equation}
Note that the cancellation of the Yukawa coupling dependence in the ratio is valid at the dominant order  in the mixing angle expansion,
 that is  to say that in Eqs.~(\ref{RmuegammaPheno})-(\ref{ratio1}) we neglected terms which involve four insertions of light-heavy mixing in the amplitudes, see Appendix. This approximation is justified for the range of low right-handed neutrino masses we contemplate, in view of the experimental
constraints on the mixing, whereas for the large mass regime it
relies on the perturbativity of the Yukawas. Note also that, for a given nucleus, the ratio may vanish for a particular value of $m_N$, see  Sec.~\ref{largemN} below.

The predictions on ratios of rates, such as that in Eq.~(\ref{ratio1}), hold as well for the case where only one right-handed neutrino would dominate the rates, because in this case the Yukawa dependence also drops, leaving the same dependence on a unique  mass.  This situation  is quite generic of scenarios where the right-handed neutrino mass spectrum is hierarchical, 
since in the large $m_N>M_W$ (low $m_N<M_W$) mass regime 
 all but the lightest (heaviest) $N_i$ contributions can be  in general neglected, see Eqs.~(\ref{asymptolargeeee}), (\ref{asymptolargegamma}), (\ref{asymptosmalleee}) and (\ref{asymptosmallgamma}).
However, the price to obtain measurable flavour-changing rates with hierarchical spectra is to induce unacceptably  large neutrino masses, disregarding eventual large fine-tuning between the various parameters~\footnote{For instance in pseudo-Dirac models as in Refs.~\cite{Gavela:2009cd,Blanchet:2009kk}, if one increases the mass splitting, one gets too large neutrino masses, unless one would have a cancellation between the different types of neutrino mass contributions, i.e.~from the Yukawa couplings whose simultaneous presence breaks lepton number and from the two right-handed neutrino mass matrix entries which break lepton number (the one contributing to neutrino masses at tree level and and the one contributing radiatively, see for example Eqs.~(3)-(9) of Ref.~\cite{Blanchet:2009kk}).} (which anyway would be destabilized by unacceptably large radiative contributions to neutrino masses~\cite{silviajacobo}). Moreover, the heavy right-handed neutrino contribution   to neutrinoless double beta decay can be in contradiction with the current bound~\cite{Blennow:2010th,Ibarra:2010xw,Ibarra:2011xn}, especially for low right-handed neutrino masses below a few tens of GeV.  In consequence, in the rest of the paper we will consider only the quasi-degenerate spectrum of right-handed neutrinos. 
 
 Although the full range of right-handed neutrino masses has been analyzed in this work, in the next sections we discuss in detail the maximum and minimum right-handed neutrino mass scales that future sensitivities may reach,  as well as the sensitivity reach for the charged-current mixing of steriles with the electrons and muon sector of the SM.  We will denote by ``large'' mass regime that in which the right-handed neutrino scale is larger than the electroweak scale, $m_N>M_W$, while the ``low'' mass regime will be that in which the heavy right-handed neutrinos are lighter than $M_W$ (although always much larger than the usual three light neutrinos).


\section {Large Mass Regime ($m_N \geq M_W$)}
\label{largemN}
Let us consider first in detail the regime of singlet fermion masses larger than the electroweak scale. 
 Before reporting on the analytical results  for $G^{\mu e}_\gamma$ and $\tilde{F}_{u,d}^{\mu e}$, it is interesting
to discuss the expected behaviour of the transition rates mediated by extra neutrino species much heavier than $M_W$. In the limit of infinite right-handed neutrino masses, the low-energy theory is renormalizable (e.g.~the Standard Model with massless left-handed neutrinos) and, in consequence, the extra singlet degrees of freedom introduced must decouple, leaving no impact on low-energy observables, see for example \cite{Grimus:2002ux}. We have checked explicitly that indeed all rates mediated by right-handed neutrinos  fulfill this condition.
 As an illustration, consider  the Z-mediated  contribution to $\tilde{F}^{\mu e}_q$ in Eq.~(\ref{BRmueexact}). Although the loop integral exhibits a logarithmic $m_N^2/M_W^2$  dependence, and furthermore longitudinal $W$s in the loop provide positive powers of $m_N$,  the $1/M$ (implying $1/m_N$) dependence of the elements of the mixing matrix -see Eq.~(\ref{approxUs})-  ensures a total rate scaling as $1/m_N^4 log^2(m_N^2/M_W^2)$  for  $m_N\gg M_W$. In other words, a rate that vanishes in the decoupling limit as it should. See also the next subsection.

\subsection{Analytical results}
 Expanding  Eqs.~(\ref{Fmue}-\ref{Fmueee}) to lowest order in inverse powers of $ x_{N_i}\equiv m_{N_i}^2/M_W^2$, the terms relevant for 
the ratio of the $\mu\rightarrow e$ conversion rate to the capture rate, Eq.~(\ref{BRmueexact}), read
\begin{eqnarray}
\tilde{F}_{u}^{\mu e}& =&\sum_{ i=1}^{k} U_{e N_i}U^*_{\mu N_i}\tilde{F}_u(x_{N_i})\,,\quad \quad \tilde{F}_u(x) = \left(\frac{2}{3}s_W^2\frac{16\log\left( x\right)-37}{12}-\frac{3+3\log\left(x\right)}{8}\right),
\label{Futilde}
\\
\tilde{F}_{d}^{\mu e}&=&\sum_{i=1}^{k} U_{e N_i}U^*_{\mu N_i} \tilde{F}_d(x_{N_i})\,, \quad \quad \tilde{F}_d(x) =  \left(-\frac{1}{3}s_W^2\frac{16\log\left(  x\right)-37}{12}-\frac{3-3\log\left( x\right)}{8}\right),
\label{Fdtilde}\\
G^{\mu e }_\gamma &=&  \sum_{i=1}^{k} U_{e N_i}U^*_{\mu N_i} G_\gamma(x_{N_i})\,,\quad \quad G_\gamma(x)=\frac{1}{2}\,,
\end{eqnarray}
whereas for the particular case of light nuclei in Eq.~(\ref{BRmueapproxlight}), it results 
 \begin{eqnarray}
F_{u}^{\mu e} &=&\sum_{i=1}^{k} U_{e N_i}U^*_{\mu N_i} F_u(x_{N_i})\,, \quad \quad {F}_u(x) =  \left(\frac{2}{3}s_W^2\frac{16\log\left( x\right)-31}{12}-\frac{3+3\log\left(x\right)}{8}\right),
\label{Fu}
\\
F_{d}^{\mu e}&=&\sum_{i=1}^{k} U_{e N_i}U^*_{\mu N_i} F_d(x_{N_i})\,, \quad \quad {F}_d(x) =  \left(-\frac{1}{3}s_W^2\frac{16\log\left( x\right)-31}{12}-\frac{3-3\log\left( x\right)}{8}\right).
\label{Fd}
\end{eqnarray}

\subsubsection*{Comparison with the literature}
Although the first two items below hold for the exact formulae and not only for the limit of large singlet fermion masses, we gather here together the result of comparing our formulae with those in previous literature (most often given in the approximation of low atomic number). A number of comments can be made:
\begin{itemize}

\item  Sign of the $G_\gamma$ contribution. Our results agree with the sign indicated in Refs.~\cite{Chang:1994hz,Ilakovac:2009jf,Deppisch:2010fr}, and is opposite to that in Refs.~\cite{Ioannisian:1999cw,Pilaftsis:2005rv}. Note that $G_\gamma$ contributes both to $\mu\rightarrow e$ conversion and to  $\mu\rightarrow e \gamma$ decay -see Eq.~(\ref{muegamma}), and the loop integrals involved can be related with analogous amplitudes in the quark sector, once the internal quark charge is switched off: this allows to check that our results are consistent with those for $K$ transitions~\cite{Inami:1980fz} and for  $b \rightarrow s l^+ l^-$ decay~\cite{Hou:1986ug}. 

\item Box diagrams.  Refs.~\cite{Chang:1994hz,Ioannisian:1999cw,Pilaftsis:2005rv} also exhibit differences in the relative size and/or sign of the crossed and not crossed box contributions, that is, in the amplitudes resulting from the last two diagrams in Fig.~\ref{diagrams}. As shown in the Appendix, we obtain a $-4$ factor between these contributions, in agreement with for instance the results for  $\Delta S=2$ or $\Delta B=2$ processes in Ref.~\cite{Buchalla:1995vs}. Also, we obtain half the contribution considered in~Ref.~\cite{AristizabalSierra:2012yy} for all box diagrams.

\item Constant terms in the $F_Z$ et $F_\gamma$ form factors.  For right-handed masses low enough to result in observable $\mu\rightarrow e$ rates,  those terms are numerically competitive with logarithmic ones and cannot be neglected, see Eqs.~(\ref{Fu})-(\ref{Fd}). We get different results for them, though, than  in Ref.~\cite{Deppisch:2010fr}, where some of them have been neglected. 
On the contrary, our results agree with the full expressions for those form factors given in in Ref.~\cite{Ilakovac:1994kj} for the computation of the $\mu\rightarrow eee$ rate, to which they also contribute. Furthermore, the $\mu \rightarrow e$ conversion rate has been also calculated in the framework of the supersymmetric type-I seesaw model  in Ref.~\cite{Ilakovac:2009jf}; we checked that its non-supersymmetric limit   results in logarithmic terms which agree with ours, but there too the constant terms have been neglected~\footnote{ Numerically, to neglect the constant terms for these form factors leads for example to a  Titanium conversion rate vanishing for $m_N\simeq 2.9$~TeV instead of 4.7~TeV,  which translates into rates  which can differ by several orders of magnitude in the few TeV range.}.

\item Decoupling limit.  As already mentioned above, in the limit in which the right-handed neutrino  masses go to infinity, Eqs.~(\ref{Fu})-(\ref{Fd}) show that these states decouple in all rates considered, as they should \cite{Grimus:2002ux}.   This is in disagreement with the non-decoupling behaviour obtained in Ref.~\cite{Dinh:2012bp} for the $\mu\rightarrow e$ conversion rate  in the  type-I seesaw model~\footnote{ The analytic expressions used in Ref.~\cite{Dinh:2012bp} are the same ones that are valid  for the case of a fourth generation of quarks and leptons~\cite{Buras:2010cp}  (that is, with an active extra neutrino instead of a singlet right-handed neutrino,  which cannot decouple as the remaining low-energy theory would not be renormalizable).}. Note that the heavy-light mixing angles are suppressed by one power of the right-handed neutrino masses, i.e.~$U\sim Y v/m_N$. In summary, in the large mass regime considered here,  $x_N=m_N^2/M_W^2\gg 1$, the leading term in the three type of rates under discussion scales as
\begin{align}
\Gamma\sim&\,(\log x_N)^2/x_N^2\, ,& & \mbox{for}\,\, \mu\rightarrow eee \quad \mbox{and}\quad \mu\rightarrow e \,\,\mbox{conversion}\,,\label{asymptolargeeee}\\
\Gamma\sim&\,1/x_N^2\, , & &  \mbox{for}\,\, \mu\rightarrow e\gamma \,.
\label{asymptolargegamma}
\end{align}

\end{itemize}

 Finally, note that with respect to the results in Ref.~\cite{Deppisch:2005zm} we get different coefficients for several logarithmic and constant terms.


\subsection{Ratios of rates involving one same flavour transition}
\begin{figure}[t]
    \centering
        \includegraphics[width=0.45\textwidth]{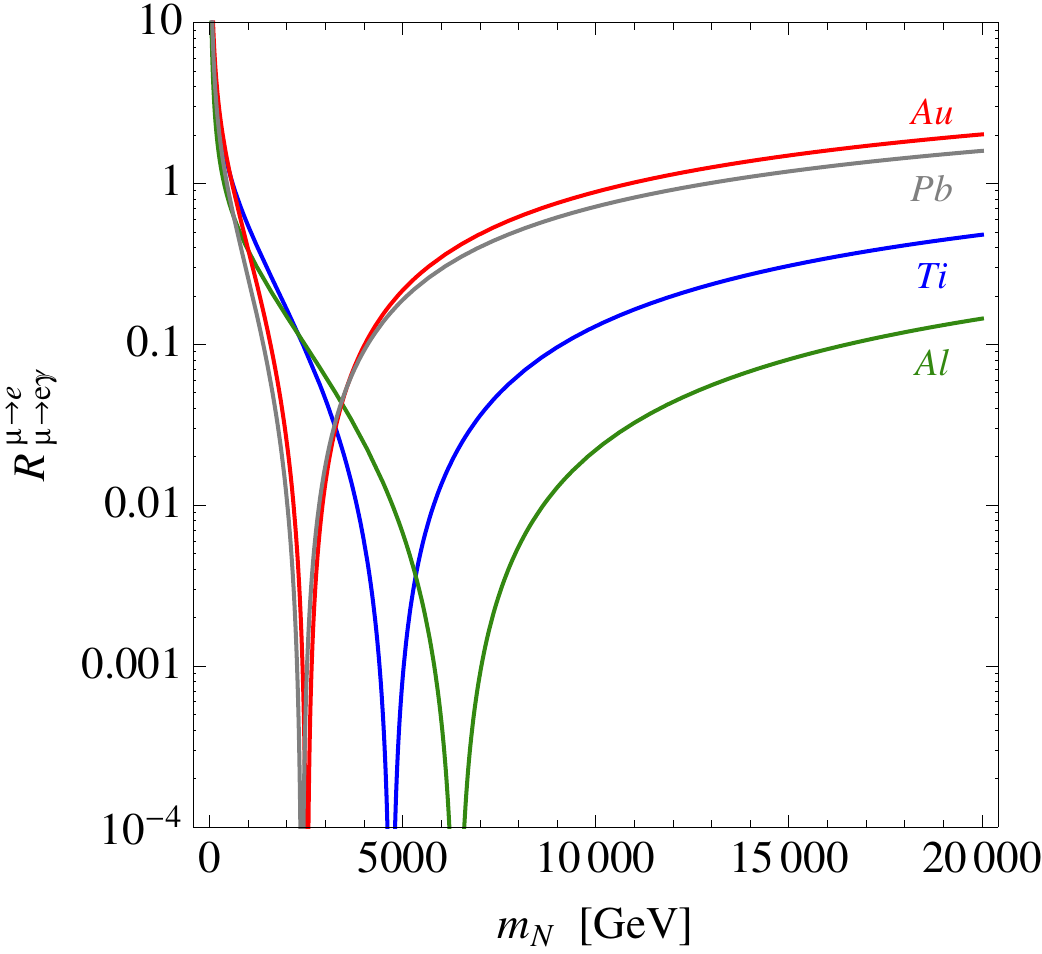}    
        \includegraphics[width=0.45\textwidth]{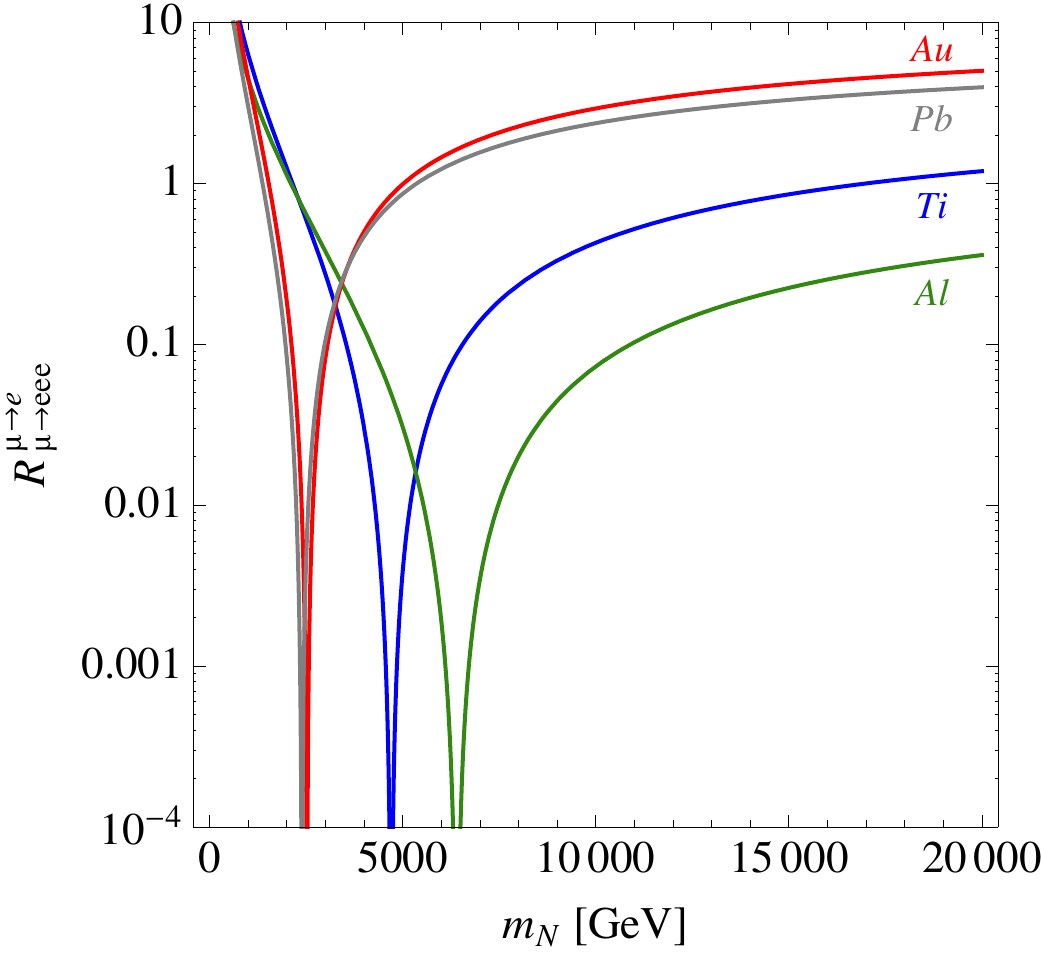} 
        \caption{ $R^{\mu\rightarrow e}_{\mu\to e\gamma}=R_{\mu\rightarrow e}/Br(\mu\to e\gamma)$ (left panel)  and $R^{\mu \rightarrow e}_{\mu\to eee}= R_{\mu\rightarrow e}/Br(\mu\to eee)$ (right panel) as a function of the right-handed neutrino mass scale $m_N$, for $\mu \rightarrow e$ conversion in various nuclei.}
        \label{Plot1}
\end{figure}

The formulae in Eq.~(\ref{BRmueexact}), (\ref{muegexactrate}) and (\ref{mueee}) allow to compare 
  the relative strength of  $\mu \to e$ conversion  to  the $\mu\to e\gamma$ branching ratio, $R^{\mu-e}_{\mu\to e\gamma}$, and to the ${\mu\to eee}$  branching ratio, $R^{\mu-e}_{\mu\to eee}$.   As explained above these ratios depend only on the right-handed neutrino mass scale $m_N$ (at leading order in $Yv/M$). The results are illustrated in  
Fig.~\ref{Plot1}  as a function of $m_N$ for various nuclei, in the large  $m_N$ regime~\footnote{For $M_{N}\approx 100$ GeV, we get $R_{\mu \rightarrow e}^{Ti}    \approx  6.7 \cdot \ Br(\mu \rightarrow e\gamma) $ and  $ Br(\mu \rightarrow eee)    \approx  0.033  \cdot Br(\mu\rightarrow  e\gamma) $, which differs with Eq. (3.21) of Ref.~\cite{Deppisch:2010fr} by two orders of magnitude.}.
One distinctive feature of these ratios is that they vanish for some value of $m_{N}$, when the $\mu \to e$ conversion rate, Eq.~(\ref{BRmueexact}), vanishes. This peculiar feature is due to the up quark $\tilde{F}^{\mu e}_u$ and down quark $\tilde{F}^{\mu e}_d$ contributions in Eq.~(\ref{BRmueexact}) having
opposite signs, 
as an outcome of their different charge and weak isospin. The precise value where the $\mu\rightarrow e$ conversion rate vanishes is nuclei-dependent and given by
 \begin{equation}
m_N^2\Big|_0=M_W^2\,\mbox{exp}\left(\frac{\frac{9}{8} V^{(n)}+\left(\frac{9}{8}+\frac{37 s_W^2}{12}\right) V^{(p)}-\frac{ s_W^2}{16 e}D}{\frac{3 }{8}V^{(n)}+\left(\frac{4 s_W^2}{3}-\frac{3}{8}\right) V^{(p)}}\right)\,,
\label{vanishmass}
\end{equation}
which shows that small variations on the nuclear form factors may result in sizeable variations on the value of  $m_N^2\Big|_0$, which is thus sensitive to the nuclear physics uncertainties. 
The uncertainty in the ratio $V^{(p)}/V^{(n)}$ translates, for instance for
$5$-$10\%$ variations, into $\cal{O}$(TeV) shifts on the value of the right-handed neutrino mass at which the conversion rate vanishes.
With the form factor values given in Table 1, the rate vanishes for mass values typically in the 2-7 TeV range, respectively 6.4, 4.7, 2.5 and 2.4 TeV for Al, Ti, Au and Pb, as  Fig.~2 shows 
~\footnote{Note that a plot of the same ratio is displayed in Ref.~\cite{Dinh:2012bp}, with quite different results, in particular vanishing rates for much lower $m_N$ values, see footnote 6 above.}.

\begin{figure}[t]
    \centering
        \includegraphics[width=0.4\textwidth]{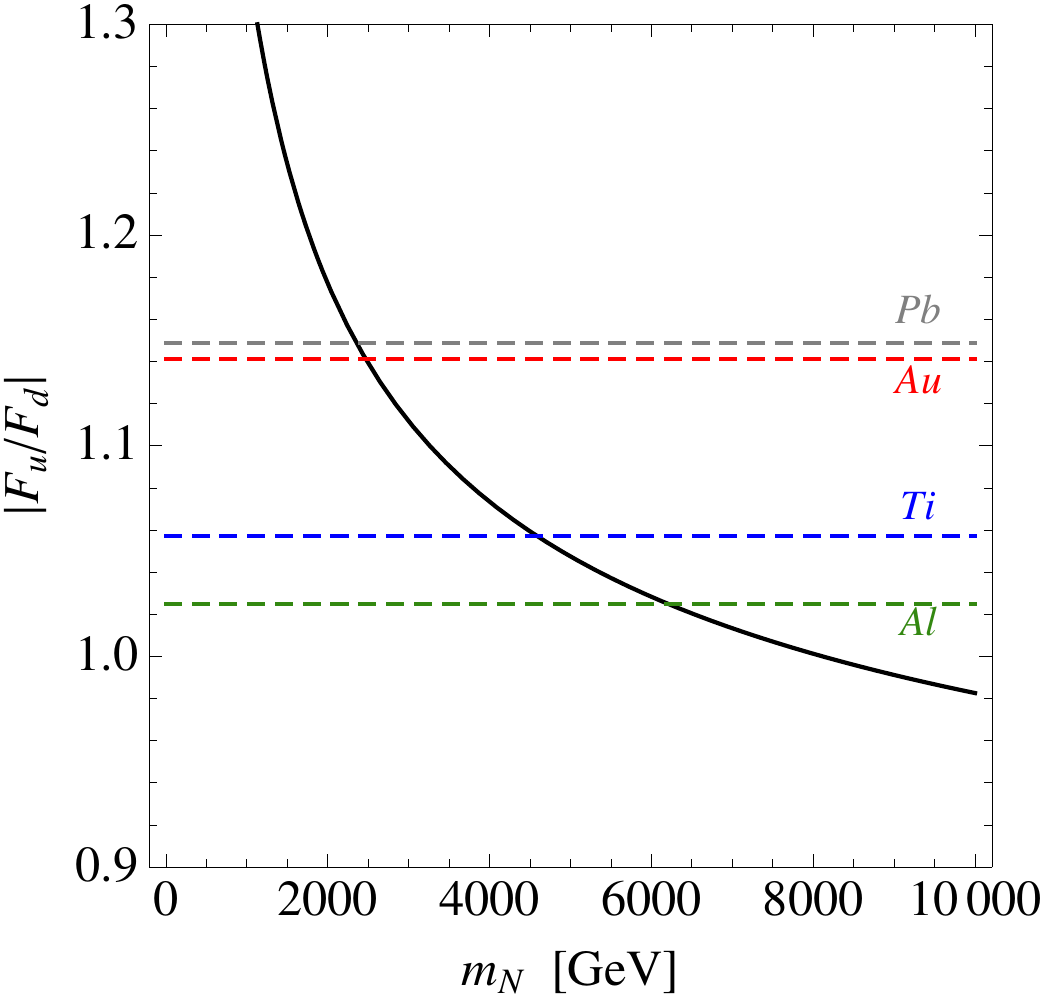}
        \caption{ $|F_u/F_d|$ (that is, $-F_u/F_d$) as a function of the right-handed neutrino mass scale $m_N$ (solid line). The dashed lines give the values of $(2A-Z)/(A+Z)$ for the various nuclei. Crossing points give the values of $m_N$ where the $\mu \to e$ conversion rate $R_{\mu\rightarrow e}$ vanishes.}
        \label{figvuvd}
\end{figure}

 For degenerate right-handed neutrinos and light nuclei, $\alpha Z \ll 1$, Eq.~(\ref{muerate}) is a good approximation which allows to rephrase the vanishing condition as 
 \begin{equation}
    \frac{F_u}{F_d} =-\frac{(2A-Z)}{(A+Z)}\,.
    \label{fufd}
\end{equation}
Due to the logarithmic behaviour of $F_u/F_d$, a small variation of ${(2A-Z)}/{(A+Z)}$ results in a sizeable variation of the value of $m_N$ for which  $R_{\mu\rightarrow e}$ vanishes. 
The atomic ratio on the right-hand side of Eq.~(\ref{fufd}) takes  the value $ -1.05$ for Ti , $ -1.02$ for Al, and  $ -1.15$ for  Au and Pb. For illustrative purposes, Fig.~\ref{figvuvd} shows the value of $F_u/F_d$  as a function of $m_N$,  together with the value of  $(2A-Z)/(A+Z)$ for each nucleus.
 Furthermore, performing in this approximation an expansion in inverse powers of $x_{N}\equiv m_N^2/M_W^2$, as given in Eqs.~(\ref{Fu}) and (\ref{Fd}), allows to rewrite the 
  mass value at which $R_{\mu\rightarrow e}$ vanishes as
\begin{equation}
m_N^2\Big|_0=M_W^2\,\mbox{exp}\left(\frac{\frac{9}{8}(A-Z)+\left(\frac{9}{8}+\frac{31 s_W^2}{12}\right) Z}{\frac{3}{8}(A-Z)+\left(\frac{4 s_W^2}{3}-\frac{3}{8}\right)Z}\right)\,.
\label{vanishmassb}
\end{equation}

\begin{figure}[h]
    \centering
        \includegraphics[width=0.4\textwidth]{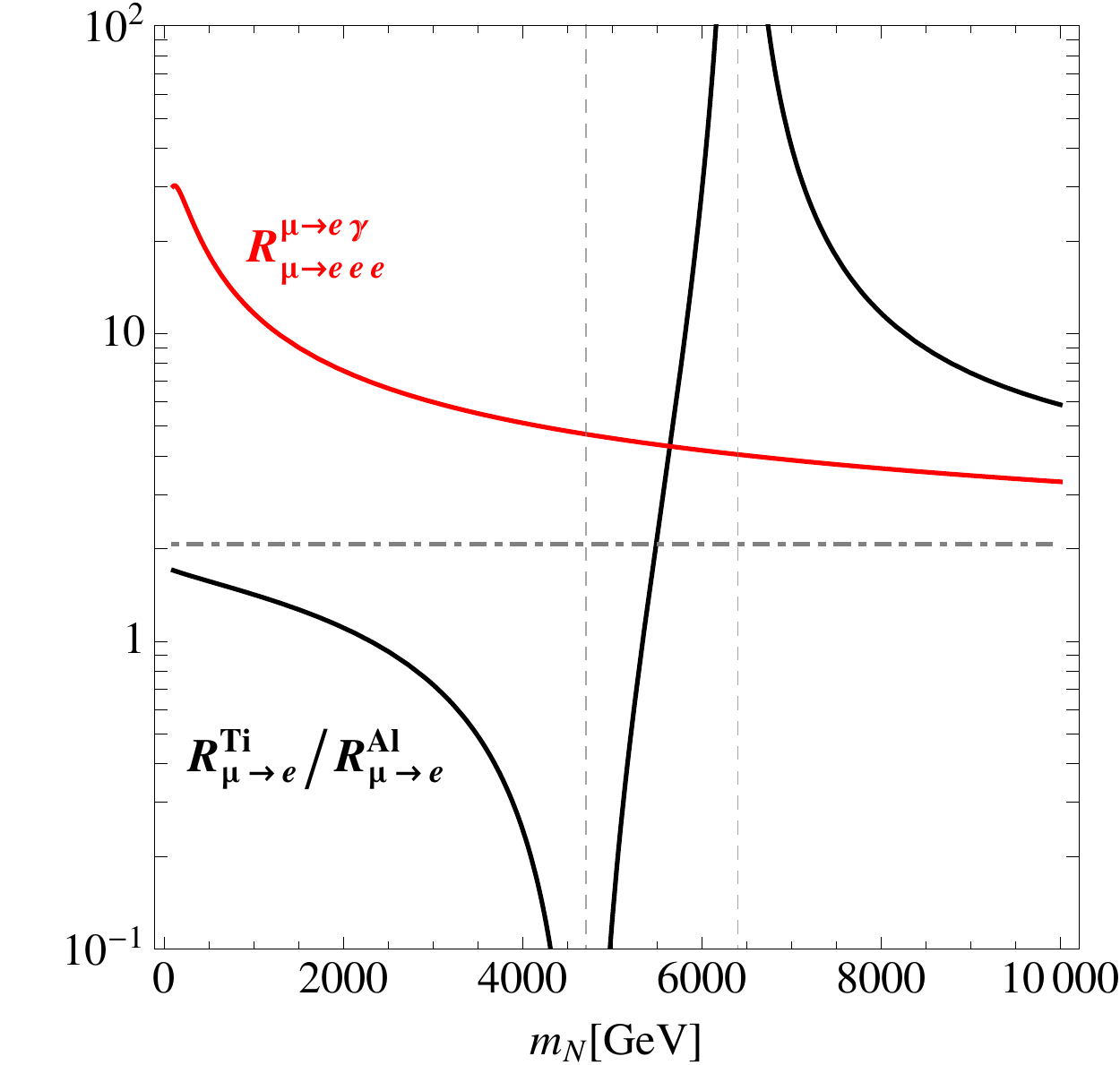} 
        \caption{$R_{\mu\rightarrow e}^{Ti}/R_{\mu\rightarrow e}^{Al}$ (black)  and $R^{\mu \rightarrow e \gamma}_{\mu\to eee}= Br(\mu \rightarrow e \gamma)/Br(\mu \rightarrow eee)$ (red) as a  function of $m_N$. The horizontal dashed line show the large $m_N$ asymptotic value of $R_{\mu\rightarrow e}^{Ti}/R_{\mu\rightarrow e}^{Al}$.}
        \label{Plot2}
\end{figure}

Fig.~\ref{Plot2} depicts  two other ratios for $\mu\rightarrow e$ transitions:  $R_{\mu\rightarrow e}^{Ti}/R_{\mu\rightarrow e}^{Al}$ and $Br(\mu\rightarrow e \gamma)/Br(\mu\rightarrow eee)$  (the latter one from Ref.~\cite{Chu:2011jg}).    $\,$Sweeping over increasing $m_N$ values, the first ratio first vanishes when $R_{\mu\rightarrow e}^{Ti}$ does, and later goes to infinity when $R_{\mu\rightarrow e}^{Al}$ vanishes.
 The second ratio, $Br(\mu\rightarrow e \gamma)/Br(\mu\rightarrow eee)$, is monotonous in $m_N$, hence it does not display any $m_N$ degeneracy.

The ensemble of the results above imply that, from the experimental determination of two $\mu-e$ transition processes, and up to discrete degeneracies, it is possible to determine the scale $m_N$ of the generic framework considered. That pair of processes could be any two among  the four processes which will be probed with improved sensitivity in near future: $\mu\rightarrow e \gamma$, $\mu\rightarrow eee$, $R_{\mu\rightarrow e}^{Al}$ and $R_{\mu\rightarrow e}^{Ti}$.

To lift possible degeneracies a third measurement may need to be considered. As an example, assume that from the MEG and COMET experiments,   $R^{\mu-e(Al)}_{\mu\to e\gamma}$  is measured to be  $\sim 0.1$: the values $m_N\approx 2.5$ TeV or  $m_N\approx 16.5$ TeV would then be singled out, see Fig.~\ref{Plot1}. To lift this degeneracy the observation of a third $\mu\rightarrow e$ transition process would be necessary: for instance $R_{\mu\rightarrow e}^{Ti}$ at PRISM or $\mu\rightarrow eee$ at $\mu3e-PSI$~\cite{Berger:2011xj}. Alternatively,  the measurement of two rates might be incompatible with the upper bound or measurement of a third one, which would rule out the scenario~\footnote{Note also that, analogously, the measurement of  $\tau\rightarrow l \gamma$ decay and of  $\tau \rightarrow l l' l'$ decay would also allow to determine the $m_N$ scale~\cite{Chu:2011jg}. That determination could be compared with the $\mu\rightarrow e$ results above, to rule out or further confirm this scenario.}. 
Similarly the measurement of a single rate, together with the upper bound or measurement of another one, could exclude this scenario for ranges of $m_N$ values (eventually excluding the whole mass range).

\begin{figure}[h]
    \centering
    \subfigure[Present bounds and future sensitivity to Yukawas as a function of
    the right-handed mass]{
        \includegraphics[width=0.4\textwidth]{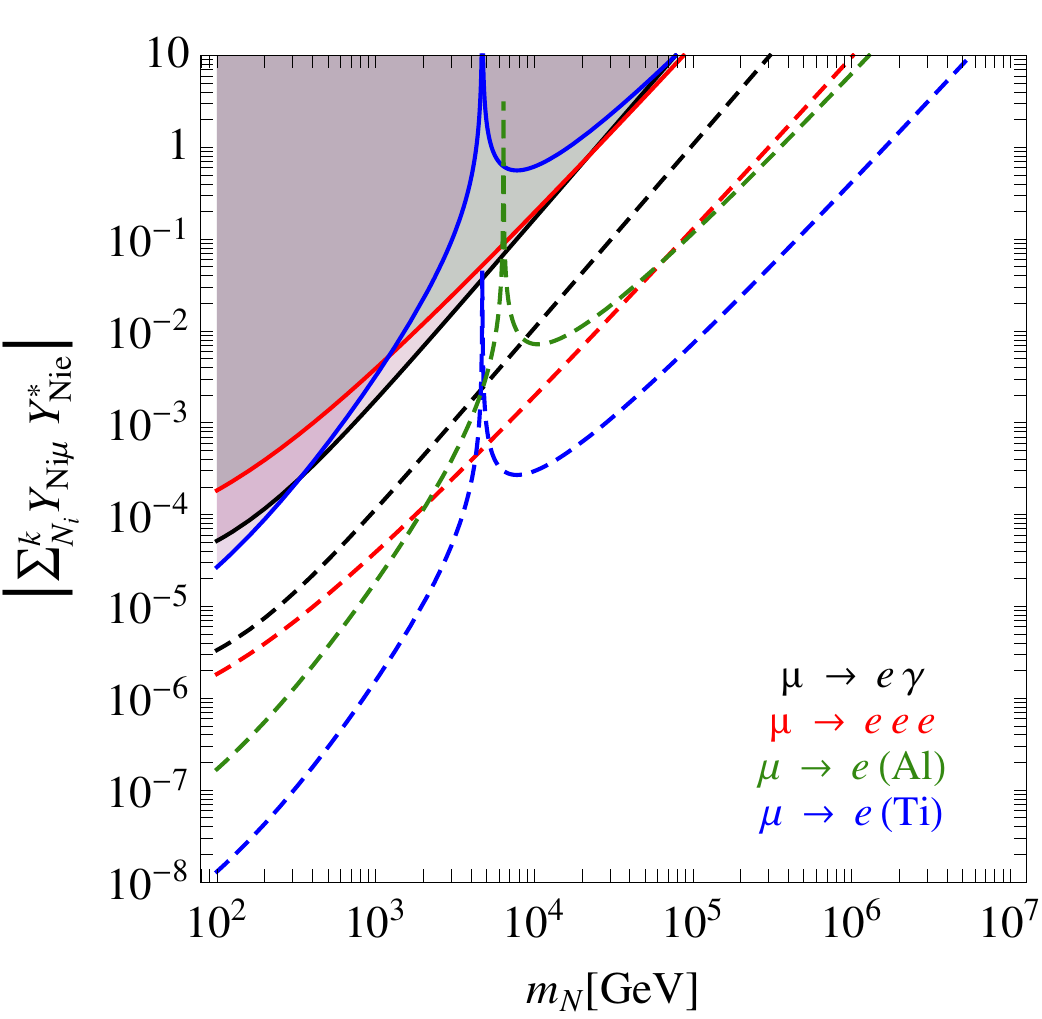} }
        \subfigure[Present bounds and future sensitivity to heavy-light mixing as a function of
    the right-handed mass]{
        \includegraphics[width=0.4\textwidth]{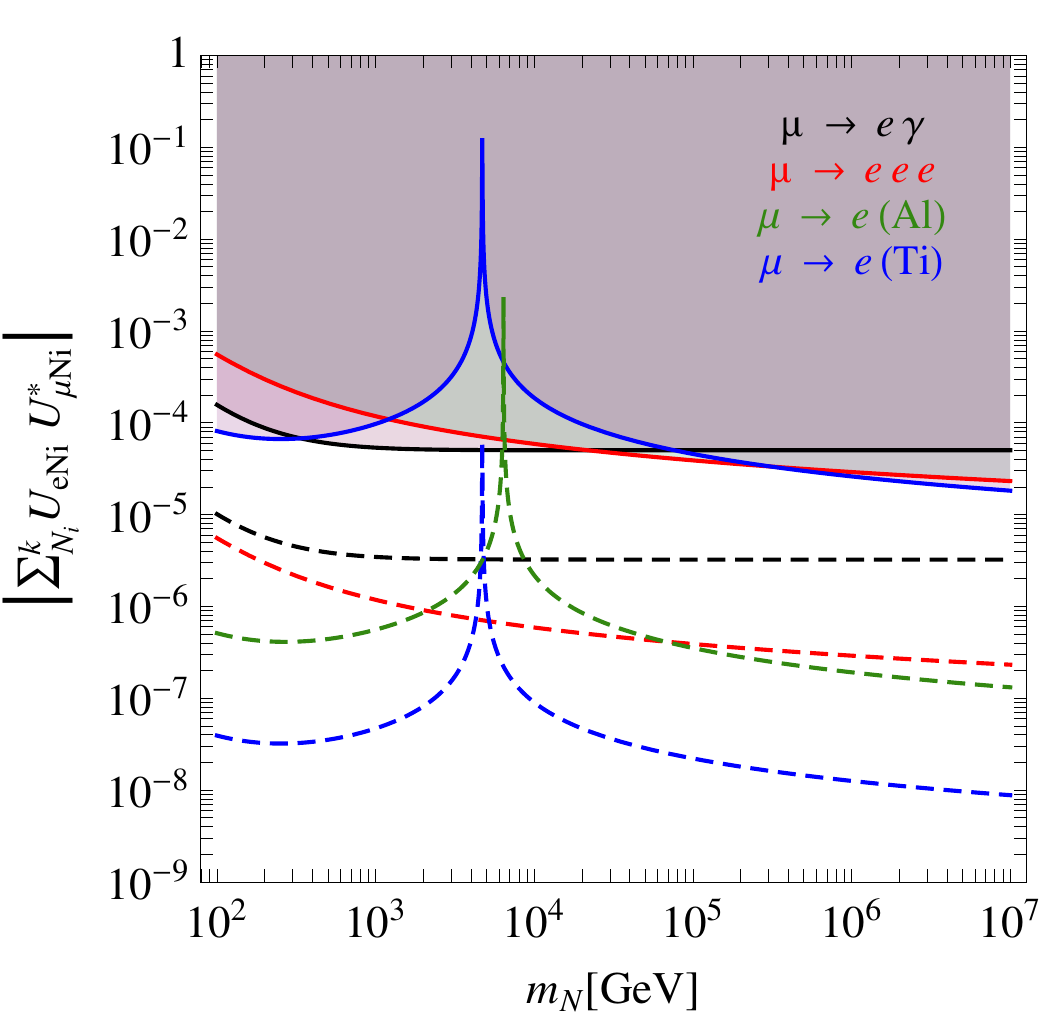} }
        \caption{Bounds on $|\sum_{i}^k Y_{N_{i\mu}}  Y_{N^*_{ie}}|$ and $|\sum_{i}^k U_{eN_i}U^*_{\mu N_i}|$ for scenarios characterized by one right-handed neutrino mass scale. The solid lines are obtained from present experimental upper bounds: from 
        Eq.~(\ref{Tipresent}) and $Br(\mu\rightarrow e \gamma)<2.4\cdot 10^{-12}$~\cite{Adam:2011ch}, $Br(\mu\rightarrow eee)<10^{-12}$~\cite{Nakamura:2010zzi}. The dashed lines are obtained from the expected experimental sensitivities: from Eqs.~(\ref{Tiexpected})-(\ref{Alexpected}) and $Br(\mu\rightarrow e \gamma)<10^{-14}$~\cite{MEG2}, $Br(\mu\rightarrow eee)<10^{-16}$~\cite{Berger:2011xj}.}       
 \label{Plot3}
\end{figure}

\subsection{Maximum seesaw scales that future experiments could probe}

 Fig.~\ref{Plot3} shows the lower bounds resulting for the Yukawa couplings and mixing parameters, if  the various rates are required to be large enough to be observed in planned experiments. It also shows the upper bounds which hold today on these quantities from the non-observation of these processes. This  figure illustrates well the impact of future $\mu \rightarrow e$ conversion measurements/bounds, as they will become increasingly dominant in exploring  flavour physics in the $\mu-e$ charged lepton sector. Values of the Yukawa couplings as low as $10^{-1}$, $10^{-3}$ and $10^{-4}$ could be probed,  for $m_N=100$~TeV, $m_N=1$~TeV and $m_N=100$~GeV, respectively, with Titanium experiments being the most sensitive.
If the Yukawa couplings are required to lie in the perturbative regime, i.e.~that each Yukawa coupling is smaller than $\sim \sqrt{4\pi}$, the bounds of Fig.~\ref{Plot3} can be rephrased as upper bounds on the $m_N$ scale:
\begin{eqnarray}
m_N&\lesssim&6000\,\hbox{TeV}\cdot \Big(\frac{10^{-18}}{R^{Ti}_{\mu\rightarrow e}}\Big)^{\frac{1}{4}}\,,\\
m_N&\lesssim&1000\,\hbox{TeV}\cdot \Big(\frac{10^{-16}}{R^{Al}_{\mu\rightarrow e}}\Big)^{\frac{1}{4}}\,,\\
m_N&\lesssim&300\,\hbox{TeV}\cdot \Big(\frac{10^{-14}}{Br(\mu\rightarrow e \gamma)}\Big)^{\frac{1}{4}}\,,\\
m_N&\lesssim&1000\,\hbox{TeV}\cdot \Big(\frac{10^{-16}}{Br(\mu\rightarrow e ee)}\Big)^{\frac{1}{4}}\,.
\end{eqnarray}
Imposing instead that the Yukawa couplings should be smaller than unity would lead to bounds smaller by about a factor of 3.
Overall, this exercise shows that future experiments may in principle probe the type-I seesaw model beyond the $\sim 1000$~TeV scale.


\section{Low Mass Regime ($m_N \leq M_W$)}
\label{lowmN}

This section focuses on the low mass 
region ($m_N\leq M_W$), to asset the discovery potential to singlet fermions  expected from future $\mu\rightarrow e$ conversion, $\mu\rightarrow e\gamma$ and $\mu \rightarrow 3e$ experiments. 
\subsection{Analytical results}
 In the low mass regime, expanding in powers of the small parameter $x_N=m_N^2/M_W^2 \ll 1$, the leading terms of the different form factors relevant for $\mu \to e$ conversion in an arbitrary nucleus -see Eq.~(\ref{BRmueexact})- are given by 
 \begin{eqnarray}
\tilde{F}_{u}^{\mu e}& =&\sum_{i=1}^{k} U_{e N_i}U^*_{\mu N_i}\tilde{F}_u(x_{N_i})\,,\quad \quad \tilde{F}_u(x) = \left(\frac{2}{3}s_W^2\frac{4\log\left( x\right)+6}{4}+\frac{3+6\log\left(x\right)}{8}\right)x\,,
\label{Futildelow}
\\
\tilde{F}_{d}^{\mu e}&=&\sum_{i=1}^{k} U_{e N_i}U^*_{\mu N_i} \tilde{F}_d(x_{N_i}) \,,\quad \quad \tilde{F}_d(x) =  \left(-\frac{1}{3}s_W^2\frac{4\log\left( x\right)+6}{4}+\frac{3}{8}\right)x\,,
\label{Fdtildelow}\\
G^{\mu e }_\gamma &=&  \sum_{i=1}^{k} U_{e N_i}U^*_{\mu N_i} G_\gamma(x_{N_i})\,,\quad \quad G_\gamma(x)=\frac{x}{4}\,,
\end{eqnarray}
whereas for conversion in light nuclei,  Eq.~(\ref{BRmueapproxlight}), they take the form 
 \begin{eqnarray}
F_{u}^{\mu e}& =&\sum_{i=1}^{k} U_{e N_i}U^*_{\mu N_i}F_u(x_{N_i})\,,\quad \quad F_u (x)= \left(\frac{2}{3}s_W^2\frac{4\log\left( x\right)+7}{4}+\frac{3+6\log\left(x\right)}{8}\right)x\,,
\label{Fulow}
\\
F_{d}^{\mu e}&=&\sum_{i=1}^{k} U_{e N_i}U^*_{\mu N_i} F_d(x_{N_i})\,, \quad \quad F_d(x) =  \left(-\frac{1}{3}s_W^2\frac{4\log\left( x\right)+7}{4}+\frac{3}{8}\right)x\,.
\label{Fdlow}
\end{eqnarray}
As a consequence, for $x_N=m_N^2/M_W^2 \ll 1$ the leading terms in the transition rates vanish as~\footnote{In the low mass regime, the amplitude for $\mu\rightarrow e\gamma$ is analogous to that for $b\rightarrow s\gamma$, while those for $\mu\rightarrow e$ conversion and $\mu\rightarrow 3e$ exhibit only a GIM cancelation quadratic in the light masses instead of the logarithmic one for quark transitions such as $b\rightarrow s e^+ e^-$, which are proportional to fermion electric charges inside the loop.}
\begin{eqnarray}\Gamma
&\sim&x_N^2(\log x_N)^2\,,\qquad \,\,\mbox{for}\,\, \mu\rightarrow eee \quad \mbox{and}\quad \mu\rightarrow e \,\,\mbox{conversion}\,,\label{asymptosmalleee}\\
\Gamma&\sim&x_N^2\,,\qquad\qquad\qquad \,\,\mbox{for}\,\,\, \mu\rightarrow e\gamma \,.
\label{asymptosmallgamma}
\end{eqnarray}
This is in contrast with the leading behaviour found for the large mass regime $x_N\gg 1$,   Eq.~(\ref{asymptolargeeee}), with the scaling law being inversely proportional to $x_N^2$, ensuring decoupling. Note that to get Eqs.~(\ref{asymptosmalleee})-(\ref{asymptosmallgamma}) we assumed fixed $U_{\nu N}$ mixing parameters, as it is customary to express constraints on sterile neutrino models in terms of these mixing parameters~\footnote{  For small sterile masses above the $eV$, the seesaw approximation  $U_{\nu N}\propto Y_N^\dagger v/m_N\,(\ll1)$ in Eq.~(\ref{approxUs}) still holds,   and an extra $x_N^{-2}$ factor has to be added in Eqs.~(\ref{asymptosmalleee})-(\ref{asymptosmallgamma}), so that the rates have a logarithmic or constant dependence on $m_N$. Below this range one enters the Dirac-dominated  regime, where the mixing angles become free parameters, and the asymptotic behaviour is that in  Eqs.~(\ref{asymptosmalleee})-(\ref{asymptosmallgamma}).}.

\subsection{Minimum seesaw scales that future experiments could probe}

\begin{figure}[h]
\centering
\subfigure[Sensitivity reach of present and future experiments in the mixing-mass parameter space]{\includegraphics[width=0.4\textwidth]{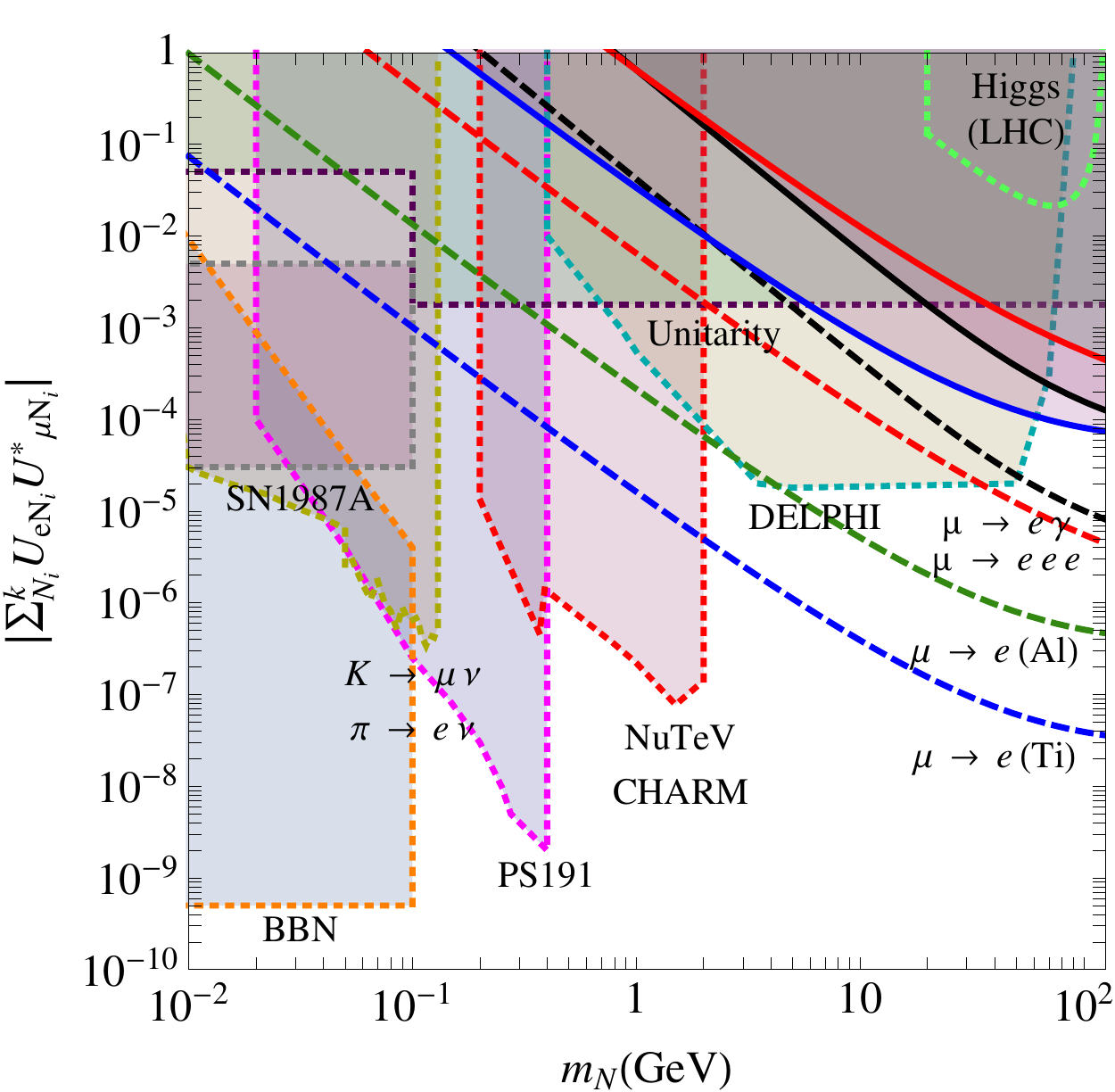}}
\hspace{5mm}
\subfigure[Maximum rates allowed by the various existing constraints]{\includegraphics[width=0.4\textwidth]{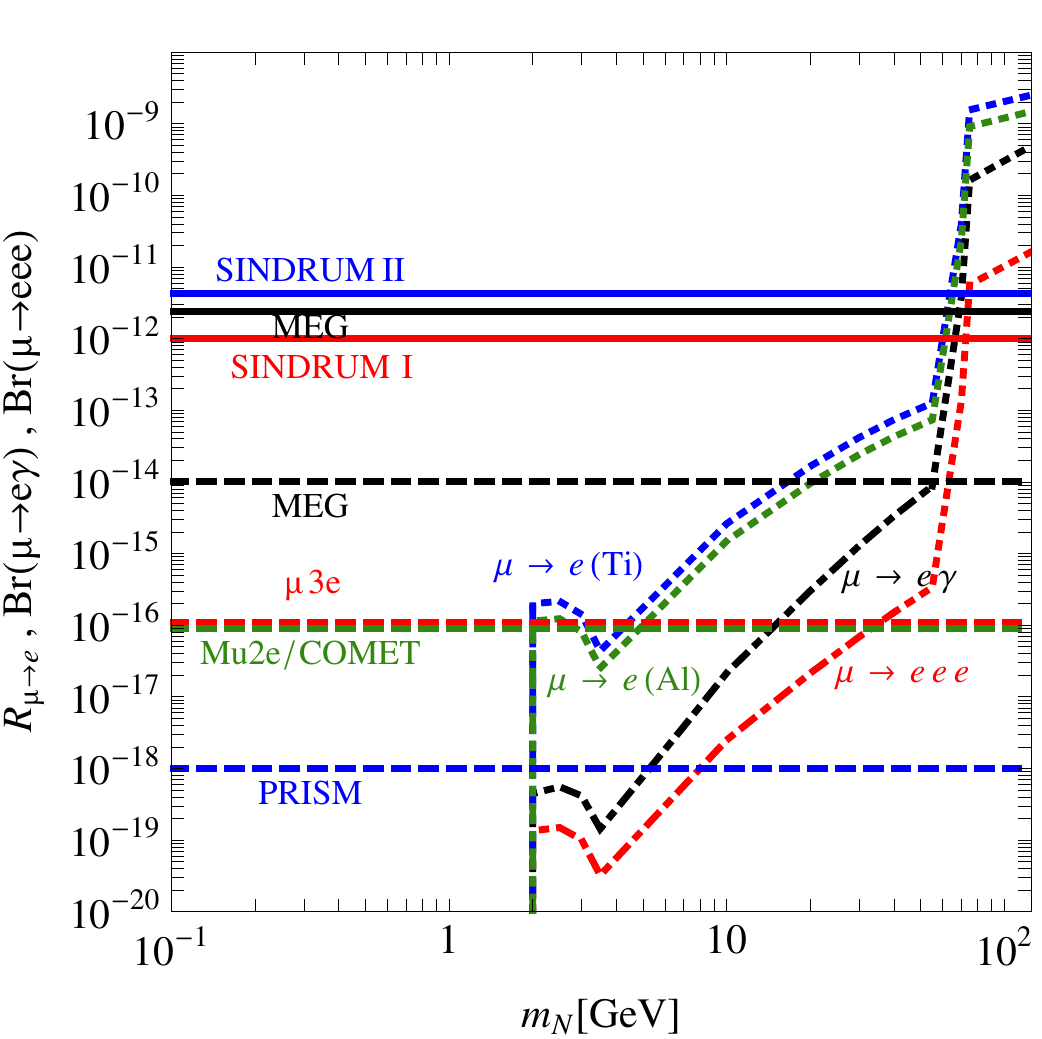}}
\caption{The left panel shows the $|\sum_i U_{e N_i}U_{\mu N_i}^*|$ versus mass sensitivity regions for present (continuous curves) and future (dashed curves)  $e-\mu$ flavour experiments. Black, red, green and
blue curves result from $Br(\mu\rightarrow e\gamma)$, $Br(\mu\rightarrow eee)$, $R_{\mu\rightarrow e}^{Al}$  and
$R_{\mu\rightarrow e}^{Ti}$, respectively. The regions already excluded 
by non-unitarity limits,   $\pi$ and $K$ peak searches, $\pi$, $K$, $D$, $Z$ decay searches, BBN, SN1987A and LHC collider searches (dotted lines) are also indicated.  Shaded areas signal the regions already excluded experimentally. 
The right panel shows the maximum allowed flavour changing rates
compatible with the bounds of the first panel. The horizontal lines give the present (solid~\cite{Dohmen:1993mp,Adam:2011ch,Bellgardt:1987du}) and future (dashed~\cite{MEG2,Berger:2011xj,Cui:2009zz,Carey:2008zz,Hungerford:2009zz}) sensitivities
of the different experiments.   }
\label{figlow}
\end{figure}

Taking $\mu \to e$ conversion experiments by themselves, the sensitivity to low singlet fermion masses is outstanding.
This is illustrated in Fig.~\ref{figlow}a (dashed lines), which shows a sensitivity down to $2$ MeV for Titanium  for large mixing.
  Nevertheless, a series of existing experimental bounds curtails the expected impact, to wit:
\begin{itemize}
\item {\it Unitarity bounds}. The mixing matrix elements $U_{\ell N}$ entering the rates are constrained by the bounds on the unitarity of the leptonic mixing matrix~\cite{Antusch:2006vwa,Antusch:2008tz}. 
 The relevant bounds here are those holding for $m_N< M_W$, in which the strong constraint from $\mu \to e \gamma$ present for masses above $M_W$ is lost due to the restoration of the GIM mechanism under the $M_W$ scale. In the region $\sim 100$ MeV$< m_N<M_W$ the bounds are dominated by the constraints on $\beta$ decay (lost under the GeV regime),  kaon decays, universality constraints from $\tau$ and $\pi$ decays, and tree-level $\mu$ decays.
 For very light $N_i$, under the pion and muon masses, the constraints from the decays of the latter are lost, since all eigenstates are then available in the decay, so that unitarity is recovered. Nevertheless, below  $100$ MeV the absence of zero distance effects in short baseline oscillation experiments (such as KARMEN~\cite{Eitel:2000by} and NOMAD~\cite{Astier:2001yj}
) also sets  constraints on the mixing elements~\cite{Antusch:2006vwa}. Stronger constraints follow nevertheless in that region, mainly from ``peak"  and ``decay" search experiments. 
  \item {\it Peak experiments}  explore the direct production of light ($<m_K,m_\pi$) extra singlet fermions in two-body ($\ell \, N_i$) particle decays of light mesons. From pion~\cite{Britton:1992xv} and kaon decays~\cite{Yamazaki:1984sj,Hayano:1982wu}, the absence of a monochromatic line -or peak-  in the charged lepton energy spectrum at $(m_{K,\pi}^2+m_\ell^2-m_{N_i}^2)/2m_{K,\pi}$~\footnote{In the rest frame 
 of the decaying meson.}, excludes at present the $30$~MeV$<m_N<400$~MeV region. Decay searches provide even stronger constraints.
\item{\it Decay experiments} including more than 2 particles in the final state look for the effects of the production and decay of massive neutrinos.  The relevant processes for constraining $\sum_i U_{e N_i}U^*_{\mu N_i}$
  are    $K,\pi,D\rightarrow \ell \,N_i\rightarrow \ell \, \ell' \nu_{\ell'} \ell''$
with $\ell ,\ell',\ell''= e,\mu $.
  Their non observation sets very strong constraints in the range $1$~MeV$<m_N<2$~GeV~\cite{Bernardi:1985ny,Bernardi:1987ek, Bergsma:1985is, Vaitaitis:1999wq}. Similarly searches for a $Z\rightarrow N_i \nu$ decay sets interesting constraints below the $M_Z$ mass~\cite{Abreu:1996pa}. For a more detailed discussion of both decay and peak experimental 
bounds see Ref.~\cite{Atre:2009rg},  and also Refs.~\cite{Ruchayskiy:2011aa} and \cite{Kuflik:2012sw}. 
\item {\it Supernovae  and BBN limits}. Upper and lower bounds on  the mixing can be obtained from supernova
SN1987A data~\cite{Kainulainen:1990bn}, since the known duration of the blast
would be modified if the right-handed neutrinos are produced in the core and escape carrying energy away. BBN limits have also been explored for the very low mass region~\cite{ Kuflik:2012sw,Kusenko:2004qc,Mangano:2011ar, Ruchayskiy:2012si}. 
  \item {\it LHC search for the decay of a 125 GeV Higgs boson via the process {\bf $\nu\, N \rightarrow \nu\, \nu \ell\ell$}}   provides new bounds on the seesaw parameters~\cite{Pilaftsis:1991ug,Chen:2010wn,Dev:2012zg,Cely:2012bz}, which can be translated into bounds on rare processes. The branching ratio of the Higgs boson decaying into Majorana neutrinos enters the implementation of these bounds,  which  in the general seesaw here considered reads 
\begin{equation}
Br( h \to \nu  N)= \frac{\alpha_W }{8M^2_W \Gamma_h^{tot} }\sum^k_i \left(| U_{e N_i}|^2+| U_{\mu N_i}|^2+| U_{\tau N_i}|^2\right) m_h m^2_{N_i }\left( 1-\frac{m^2_{N_i }}{m^2_h}\right)^2\,,
\label{Higgs}
\end{equation}  
where the decay into all three light neutrinos $\nu_j$ and all heavy ones $N_i$ is considered. $\Gamma_h^{tot}$ denotes the total decay width of the Higgs boson including  the SM channels plus those producing the extra heavy neutrinos. 
The different decay widths of the heavy neutrinos also enter the analysis, mainly all $N\rightarrow \nu\ell^+ \ell^-$ channels.
   \end{itemize}
 Following Ref.~\cite{Atre:2009rg} (see also Refs.~\cite{Ruchayskiy:2011aa,Kuflik:2012sw,Ruchayskiy:2012si,de Gouvea:2007uz}), Fig.~\ref{figlow}a  provides an approximate drawing of  the regions  excluded by these constraints (depicted as shaded areas),  in particular non-unitarity bounds, kaon and pion peak searches, kaon decay searches by the PS191 experiment, $D$ meson decay searches by the  CHARM~\cite{Bergsma:1985is} and  NuTeV~\cite{Vaitaitis:1999wq} experiments~\footnote{  
 The bounds from meson decay searches for masses below $2$ GeV depicted in Fig.~\ref{figlow} 
 are those provided by the experimental collaborations which assume decaying Dirac neutrinos; for Majorana neutrinos they may differ by factors $\sim \sqrt{2}$ or less, depending on whether the decay channel is or is not self-conjugate~\cite{Bernardi:1985ny}. The inclusion of those factors would require a reanalysis of the experimental data for masses lighter than $2$ GeV, which we refrain from attempting here as that region turns out to be out of reach for the $\mu\rightarrow e$ conversion experiments under discussion.}, $Z$ decay searches by the Delphi experiment~\cite{Abreu:1996pa}, and Higgs decays data from LHC. The exclusion lines for the mixing parameter $| \Sigma _i U_{e N_i} U^*_{\mu N_i }|$ are valid at $90\%$ CL, for the present (continuous curves) and future (dashed curves) reach of the three types of measurements under discussion: $\mu \rightarrow e\gamma$,
$\mu \rightarrow eee$ and $\mu \rightarrow e$ conversion.  Note the impressive sensitivity expected from
$\mu\rightarrow e$ conversion in Titanium taken by itself, reaching seesaw masses down to $2$ MeV.  
    Nevertheless, the very  stringent PS191, CHARM and NuTeV bounds on decay searches into $N_i$ determine a lower bound
on the mass at which conversion experiments would be competitive:  $2$~GeV.
  
Fig.~\ref{figlow}b shows 
the maximum rates allowed by the bounds on mixing depicted in 
Fig.~\ref{figlow}a (shaded areas), 
 as a function of the mass $m_N$, at the $90\%$ CL.  
Horizontal lines in Fig.~\ref{figlow}b indicate the present (continuous) and future (dashed)
sensitivities of the different experiments: their intersection with the corresponding rate (depicted with the same color) determines the lowest mass for which the new experiments would improve present bounds.

\begin{figure}[t]
     \centering
 \subfigure{\includegraphics[width=0.4\textwidth]{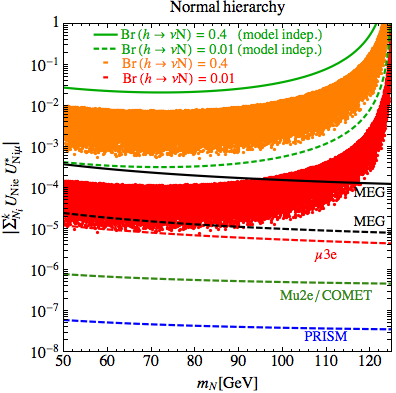}}
 \subfigure{\includegraphics[width=0.4\textwidth]{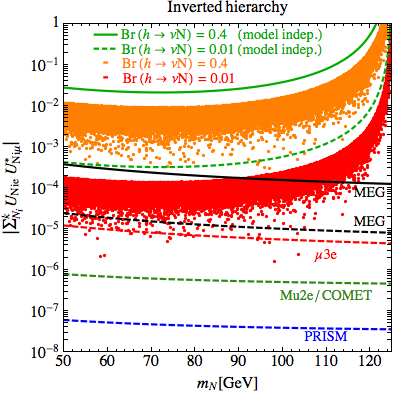}}
  \caption{Bounds on $| \Sigma _i U_{e N_i} U^*_{\mu N_i}|$  from LHC Higgs decay data and comparison with flavour violation searches. 
   The contraints from LHC data are illustrated for a sensitivity of $Br(h\rightarrow\nu N) < 0.4$, and a future one improved to the \% level~\cite{Cely:2012bz}.  For these values,
   isolated green curves are the absolute bounds  for a generic seesaw model. Bands in orange and red show instead the variation with the unknown values of the  Dirac and Majorana CP phases, for the approximately L conserving scenario in Ref.~\cite{Gavela:2009cd}, for normal (lelft panel) and inverted (right panel) hierarchy. The present (black) and future (dashed black) MEG sensitivities and  the expected one for conversion in Titanium (blue), in Aluminium (green) and $\mu\rightarrow eee$ (red) are shown for comparison.}
        \label{PlotHiggs}
\end{figure}

About the Higgs boson decay constraints, Fig.~\ref{figlow}a depicts a model-independent bound, except for the assumption of degenerate heavy neutrinos (or hierarchical ones, e.g. dominated by only one heavy species).  This bound is obtained~\cite{Dev:2012zg} by 
translating the absence of an excess over the SM expectation  for the channel $h\rightarrow \nu\bar\nu \ell^+\ell^-$ into an upper bound on Br$(h\rightarrow \nu N)< 0.4$~\cite{Espinosa:2012im} in Eq.~\ref{Higgs},  and employing then the general inequality $| \Sigma _i U_{e N_i} U^*_{\mu N_i }|<\Sigma_{i,\alpha} |U_{\alpha N_i}|^2$
~\footnote{In Fig.~\ref{figlow}a this bound is plotted down to 20 GeV, a point below which the typical cuts in the invariant mass of the lepton pair would remove the events coming from  higgs decay via the right-handed neutrino~\cite{Dev:2012zg}. The equivalent bounds shown in \cite{Dev:2012zg,Cely:2012bz}, which we used, do not extend further down than $50$ GeV: in the region $20-50$ GeV  what Fig.~\ref{figlow}a depicts is an extrapolation. }.
Figs.~~\ref{figlow}a and \ref{PlotHiggs} illustrate that it turns out to be less stringent than for example the present MEG bound on $Br(\mu\rightarrow e \gamma)$. In the future, if LHC can reach a  $\sim 1\%$ sensitivity on Br$(h\rightarrow \nu N)$, the constraint would be comparable to the present MEG one.

 More stringent bounds follow in concrete realizations of the seesaw. For instance,  the scenario with just two right-handed neutrinos added to the SM and approximate L conservation in Ref.~\cite{Gavela:2009cd,Blanchet:2009kk} (see also Refs.~\cite{Barbieri:2003qd,Raidal:2004vt,Kersten:2007vk,Asaka:2008}) is very predictive. The mixing-dependence of Br$(h\rightarrow \nu N)$ -the term between parenthesis in Eq.~\ref{Higgs}-  reduces in this case to an overall scale dependence, which can then be bounded from Fig.~\ref{figlow}a.    
 Furthermore, in these scenarios   the mixing elements
  $|U_{\alpha\,N_i}|$ are not arbitrary, but explicit functions of a few observable quantities such as the measured light neutrino mass differences and mixing angles,  the overall scale and the Dirac ($\delta$) and Majorana ($\alpha$) phases. This allows to express bounds on   $|U_{eN_i}U_{\mu N_i}^*|$ from Higgs decay as a function of the values of those CP phases. This interesting fact is illustrated in Fig.~\ref{PlotHiggs}, whose bands depict the maximum and minimum bounds obtained by varying those  phases.  The values of the neutrino parameters used in this figure are  $\Delta m_{21}^2  =(7.59 \pm 0.2)\cdot 10^{-5}$ eV$^2$, 
 $|\Delta m_{31}^2| =(2.36 \pm 0.2)\cdot 10^{-3}$ eV$^2$ for the case of neutrino inverted hierarchy (IH), 
 $|\Delta m_{31}^2| =(2.46 \pm 0.12) \cdot10^{-3}$ eV$^2$ for the case of neutrino normal hierarchy (NH), 
 $\theta_{12} =34.4 \pm 1$ degrees, $\theta_{23} =42.8^{+4.7}_{-2.9}$ degrees, 
and $\sin^2 2 \theta_{13}=0.092\pm0.017$ ~\cite{GonzalezGarcia:2010er,An:2012eh}. For NH, the maximum and minimum boundary lines  of the bands  correspond approximately to 
 $(\delta,\alpha=  0, \pi/2)$  and $(\delta,\alpha=  0, -\pi/2)$, respectively, while the equivalent values for the case of IH are 
    $(\delta,\alpha= 3\pi/2, \pi)$  and $(\delta,\alpha=  0, -\pi/4)$. 
   The values allowed never reach the absolute bound, also depicted.  On the other side, note that the plot for IH allows some points in which the $\mu-e$ mixing tends to vanish: this is expected and well-known (first pointed out in Refs.~\cite{Alonso:2010wu,dePablo:2010zz}), because in this class of models some entries of the Yukawa couplings may vanish (and therefore the entry in the mixing $U_{\alpha N_i}$ would vanish too) for certain values of the CP phases and light neutrino mixing and mass differences within their 90\% C.L. region \cite{Chu:2011jg}.
    As an example of the impact of the  analysis for this type of models, with the present limit on Br$(h\rightarrow \nu N)$, if MEG observes a signal of order the present sensitivity, only particular ranges of the phases would be allowed in this model.  In any case, for the $\mu$-e channel it is not expected that LHC data will allow to approach, even from far, the sensitivities one expects in the new generation of $\mu\rightarrow e$ conversion experiments, Eqs.~(\ref{Tiexpected}) and (\ref{Alexpected}).

\begin{figure}[h]
\centering
\subfigure[Ratio of the conversion rate to $\mu\rightarrow e\gamma$]{
\includegraphics[width=0.4\textwidth]{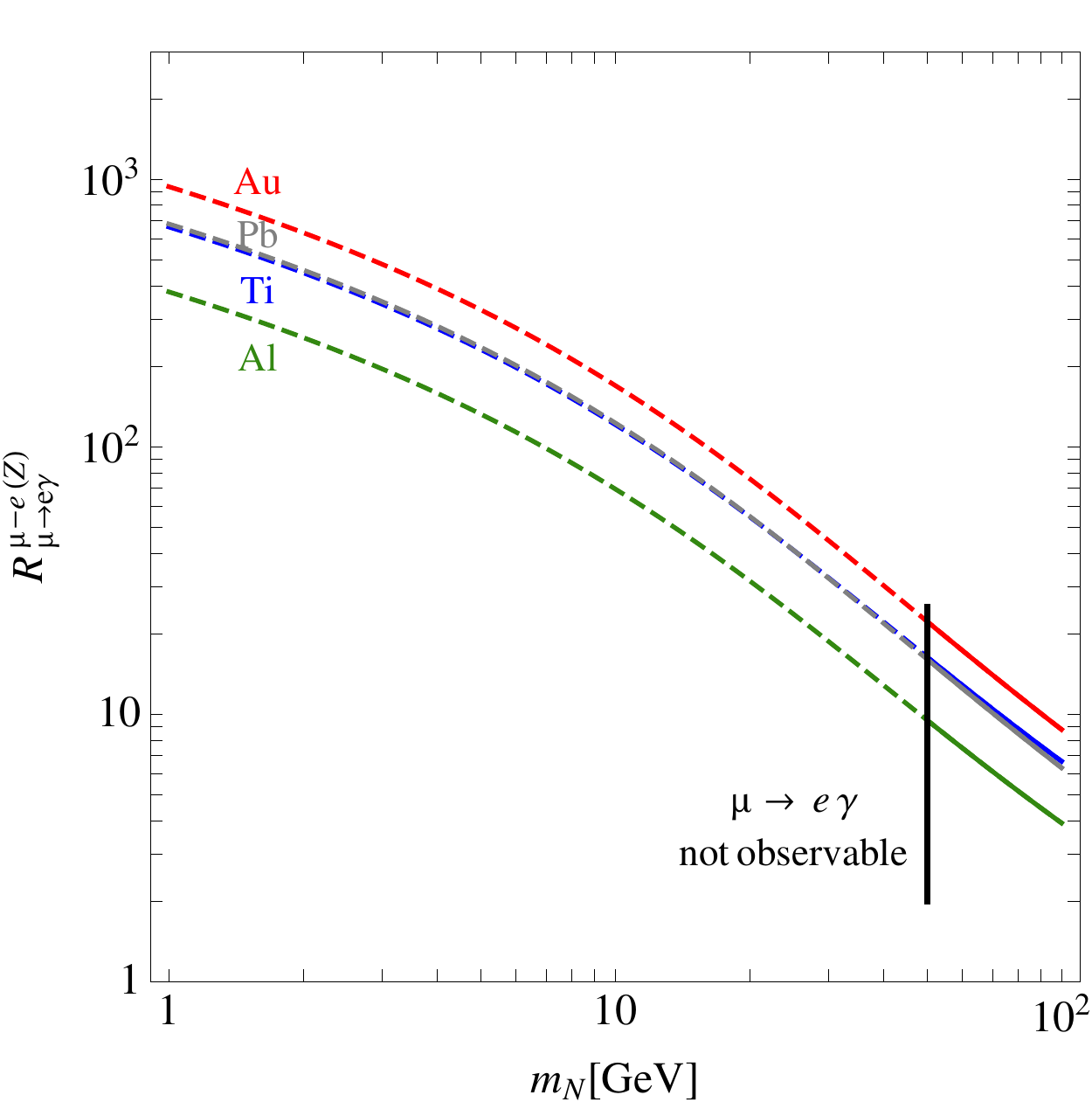}}
\subfigure[Ratio of the conversion rate to $\mu\rightarrow eee$]{
\includegraphics[width=0.4\textwidth]{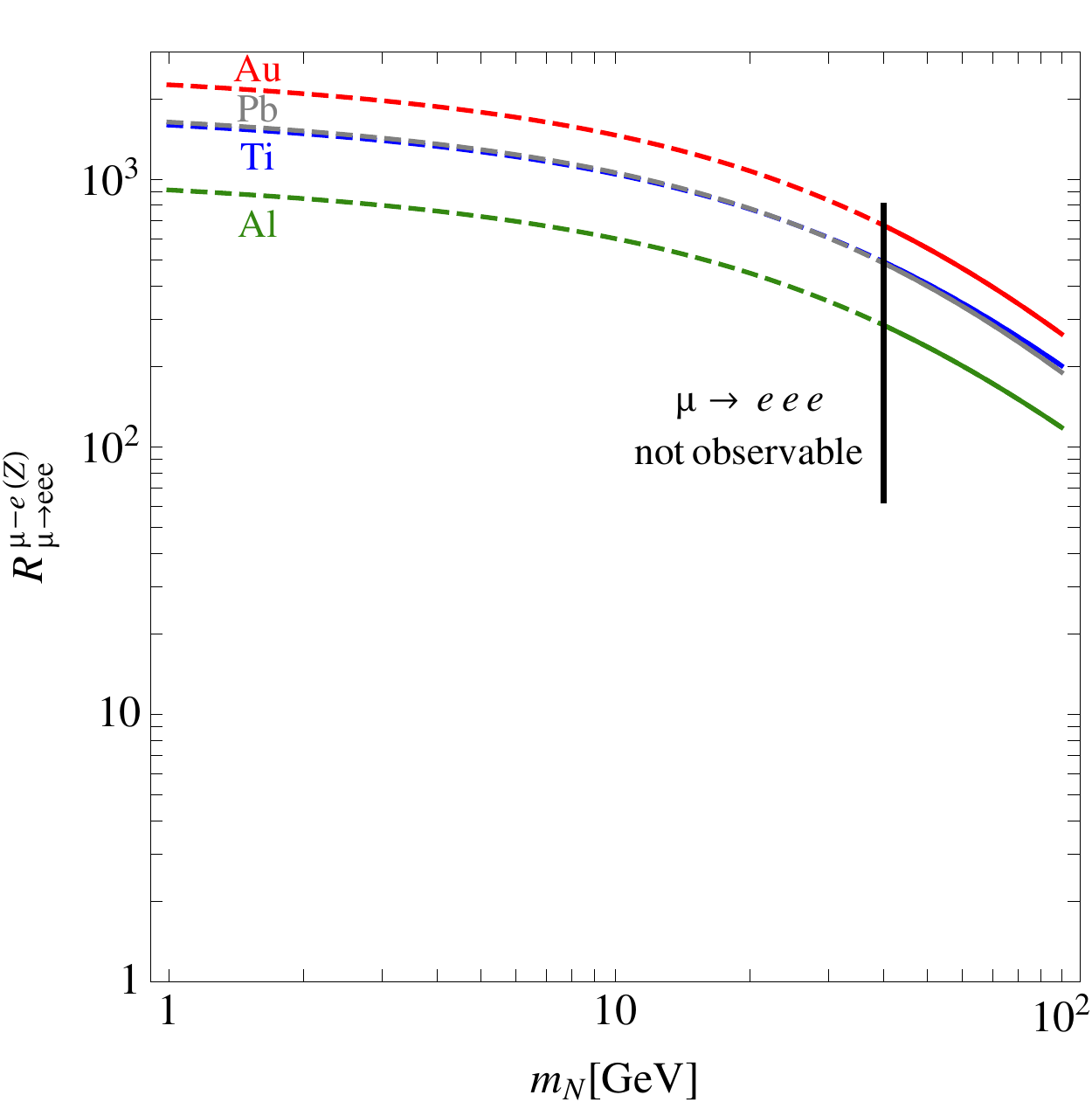}}
\caption{$\mu\rightarrow e$ conversion rate over $Br(\mu\to e\gamma)$ (left panel)  and  over $Br(\mu\rightarrow eee)$ (right panel) as a function of the right-handed neutrino mass scale $m_N$, in the light regime ($m_N<M_W$) for various nuclei. Curves become dotted lines at the
limit of observability by future experiments.}
\label{figratiolow}
\end{figure}

Finally,  Fig.~\ref{figratiolow} depicts the ratio of $\mu\rightarrow e$ conversion rate to $Br(\mu\rightarrow e \gamma)$ and  to $Br(\mu\rightarrow eee)$, in the low mass region of the type-I seesaw scenario considered. These ratios are now always larger than one,  and they do not vanish for any value of the singlet fermion mass, in contrast with the behaviour in the large mass regime discussed in Sect.~\ref{largemN}.
 With diminishing mass, the ratio $R^{\mu\rightarrow e }_{\mu \rightarrow eee}$ becomes constant, while the ratio $R^{\mu\rightarrow e}_{\mu \rightarrow e\gamma}$
grows logarithmically, see   Eqs.~(\ref{Futildelow})-(\ref{Fdlow}). The dotted section of the curves indicates the mass region which is out of observability reach for the planned experiments under discussion. 

Note that in Fig.~\ref{figratiolow}   the ratios to $\mu\rightarrow e\gamma$
and $\mu\rightarrow eee$ for Titanium and Lead fall on top of each other. Numerically the agreement holds at the $\%$ level. 
 This is in fact due to 
 the conversion ratio $R_{\mu e}$ in both nuclei being approximately equal, 
at the percent level  (modulo possible nuclear  physics  uncertainties).
 The ratios entering this quantity vary 
  from atom to atom solely through the variables $V^{(p)}/\sqrt{\Gamma_{capt}}\,,\,V^{(n)}/\sqrt{\Gamma_{capt}}\,,\,D/\sqrt{\Gamma_{capt}}$, see Eq.~(\ref{BRmueexact}), which  differ by about $10\%$ for Lead versus Titanium, and similarly for Gold for instance, see Table~1. But the combination in which these variables enter in Lead and Titanium, via the form factors $\tilde{F}_{u,d}^{\mu e}\,,G_\gamma^{\mu e}$
in Eq.~\ref{BRmueexact}, happens to be very close numerically. The
coincidence at the \% level of the $\mu \rightarrow e$ conversion ratios $R_{\mu e}$
in Titanium and Lead in all the mass range under $50$ GeV is a prediction of the quasi-degenerate right-handed neutrino scenario for very low seesaw scales.

It is worth to remark that the sensitivities and exclusion plots obtained apply not only to the (quasi-degenerate) type-I seesaw scenario object of this work, but to any BSM renormalizable theory containing Dirac or pseudo-Dirac singlet fermions, which mix with SM light leptons with strength $U_{\ell N}$.

\section{Conclusions}
\label{conclusions}

Future experiments aiming to detect  $\mu\rightarrow e$ conversion in atomic nuclei are especially promising for the discovery of  flavour violation in charged-lepton transitions.
For various nuclei, we have calculated the $\mu\rightarrow e$ rate which holds in the framework of the type-I seesaw scenario of neutrino masses.
We discussed the phenomenological impact and the reach of present and planned experiments 
in terms of the seesaw scale and mixing. 

Analytically, the $\mu \to e$ conversion rate includes form factors with and without a logarithmic dependence on the heavy singlet fermion masses. For the former (unlike for the latter) our results agree with those in Ref.~\cite{Deppisch:2010fr} (and in Ref.~\cite{Ilakovac:2009jf} provided one takes the non-supersymmetric limit of the results quoted there). We basically disagree at different levels with all other calculations we have found in the literature, be it for the constant or the logarithmic terms. The constant terms in the form factors turn out to be numerically competitive with the logarithmic ones and must be taken into account. 

The computation performed is valid for all values of the heavy right-handed neutrino masses above the MeV scale (that is, sizeably larger than the masses of the three known neutrinos), and the complete expressions have been used to obtain the numerical results. 
 Furthermore, for illustrative purposes, the leading behaviour in the regime of large ($>M_W$) and of small ($<M_W$) right-handed neutrino mass scales has been explicitly discussed.  
 
\begin{figure}[h]
\centering
\includegraphics[width=0.6\textwidth]{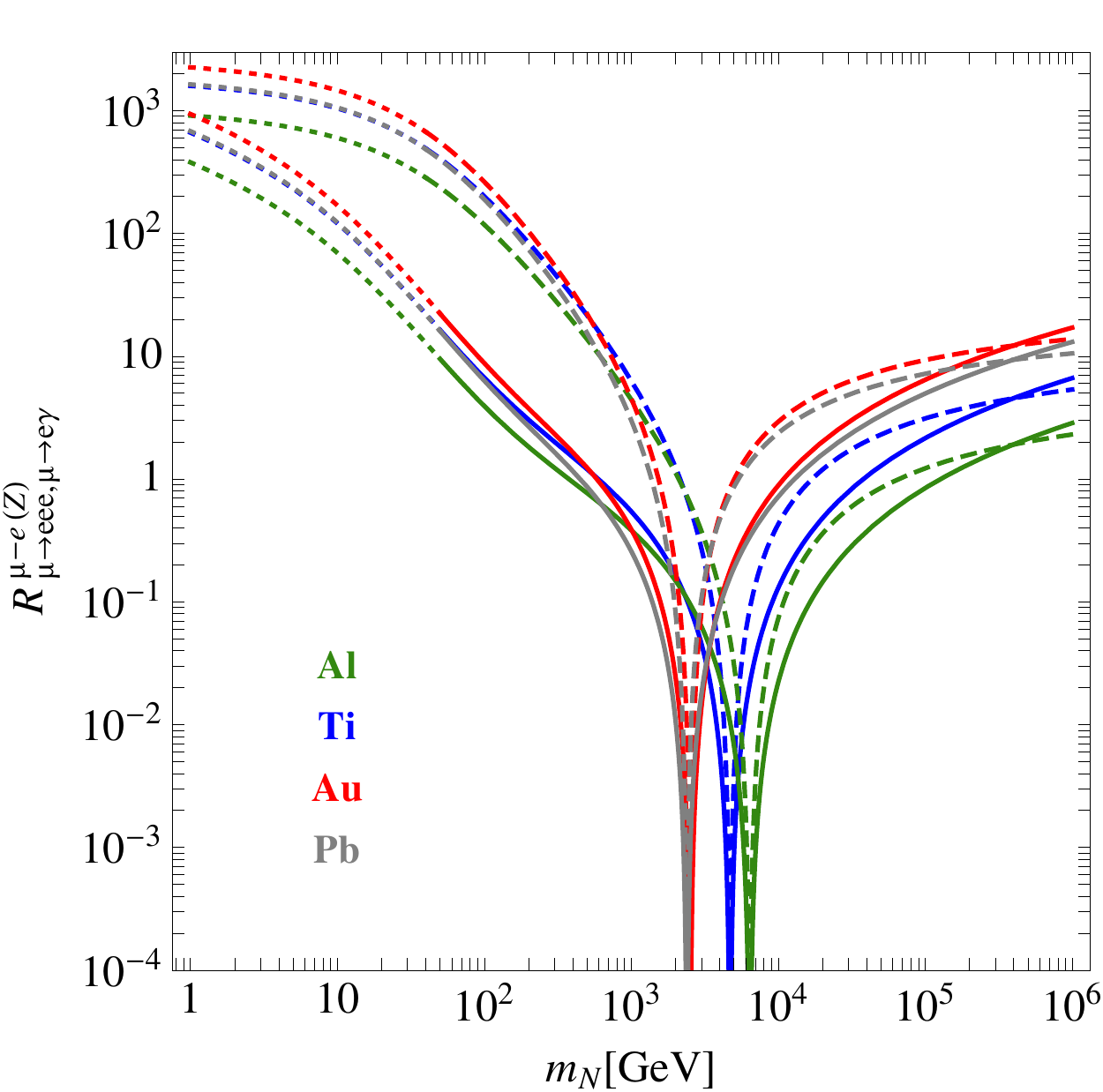}
\caption{Ratio of the $\mu\rightarrow e$ conversion rate in Al (green lines), Ti (blue lines) and Au (red lines) to  $Br(\mu\rightarrow e \gamma)$ (solid lines) and to $Br(\mu\rightarrow eee)$ (dashed lines) for the entire range in mass $m_N$
here considered. Lines are dotted when they require, for $\mu\rightarrow e \gamma$ and $\mu\rightarrow eee$, a sensitivity better than the one expected at planned experiments.}
\label{figratiocomplete}
\end{figure}

 The various possible ratios of rates  involving the same charged $\mu-e$ flavour transition, have been determined.  When a single heavy scale dominates, they depend exclusively on that unique  scale. 
    The results  are illustrated in Figs.~\ref{Plot1} and \ref{Plot2} for the large mass regime, and  in Fig.~7 for the low mass regime. 
 Fig.~\ref{figratiocomplete} provides a summary of these results by merging these figures together.
It is useful for comparison purposes and illustration of degeneracies within one same experiment and across experiments, especially if a positive signal is observed.
These results apply both in the limit of non-degenerate  heavy neutrino masses and in the hierarchical limit.  We focused the discussion on the quasi-degenerate scenario, as it is the most natural one to allow for observable rates.

 Since the  various ratios exhibit a different mass dependence, they offer a particularly neat way to confirm/exclude  possible scenarios. In the case of an agreement between theory and  measurements, they would allow a determination of the right-handed neutrino mass scale. For right-handed neutrinos in the large mass regime, depending on the ratio which would be measured, one may have a degeneracy in the right-handed neutrino mass scale it could correspond to. This can be resolved by the measurement of another ratio. Furthermore, we find that the values of the heavy neutrino masses for which the ratios of $\mu \to e$  conversion to  $\mu$ decays may vanish  are around $2-7$ TeV.
  
  We have explored the sensitivity to the singlet fermion mass scale as a function of the mixing with the muon and electron sector of the SM. The maximum scale that future $\mu \to e$   conversion experiments could probe is above the $1000$ TeV scale. This sensitivity extends to very low masses, down to $\sim2$ MeV for Titanium. When the stringent bounds coming from massive sterile neutrino searches (from the unitarity limits of the $U_{PMNS}$ mixing matrix,  from $\pi$, $K$ and $D$ meson decay searches, and other constraints) are taken into account, the expected impact in constraining the low mass regime is somewhat reduced. We have discussed as well the LHC sensitivity from Higgs decay via heavy neutrinos, both in a model-independent way and for the case of seesaw models with approximate lepton number symmetry, comparing it with present and future  $\mu-e$ flavour experiments. 
Furthermore, an interesting prediction  is the coincidence at the percent level of the conversion rate for Titanium and Lead experiments,  in the full range of  seesaw masses under $50$ GeV.
  
   Considering our results and all constraints together, it follows that planned  $\mu \to e$ conversion experiments will be fully relevant to detect or constrain sterile neutrino scenarios in an impressive mass range of $2$ GeV - $1000$ TeV. Note that this result is of general interest, as it may apply to any renormalizable BSM or dark matter theory which contains singlet fermions in its spectrum.

\section*{Acknowledgements}

We thank E. Fernandez-Martinez, P. Hernandez, S. Pascoli and  A. Poves for useful discussions. R. A. and M.B.G. acknowledge partial support
from the European Union FP7 ITN INVISIBLES
(Marie Curie Actions, PITN- GA-2011- 289442), from the CiCYT
through the project FPA2009-09017 and from CAM through the project HEPHACOS P-ESP-
00346. R. A. acknowledges MICINN support
through the grant BES-2010-037869. The work of M.D. is supported by a FRIA Ph.D.~grant. 
M.D.~and T.H.~acknowledge support from the FNRS-FRS, the IISN and the Belgian Science Policy (IAP VI-11). T.H. thanks the Departamento de F\'isica Te\'orica (UAM-Madrid) and the IFT-Madrid for hospitality and the Universidad Aut\'onoma de Madrid and CAM (Proyecto HEPHACOS S2009/ESP-1473) for partial support. Finally, we thank the Galileo Galilei Institute for Theoretical Physics for their hospitality and the INFN for partial support during the completion of this work.
\appendix

\section{Appendix: amplitudes}

As discussed in Section~\ref{muegammageneral}, the calculation of the $\mu\rightarrow e$ conversion rate requires the separation of the local  and the long-range contributions. The latter is given by the dipole term in Eq.~(\ref{muegamma}), 
while the local contribution is determined from  $Z$ and $W$ mediated diagrams, as well as from the photon monopole interaction
with the leptons. The amplitudes associated to these three ingredients can be written as~\footnote{The term proportional to $q_\mu$ in Eq.~(\ref{muegamma}) drops because of quark current conservation.}
\begin{eqnarray}
	i\mathcal{M}^q_{\gamma-monopole} &=&  \frac{i\alpha_W \alpha }{2 M^2_W} \overline{u}_q  \gamma^\mu  \mathcal{Q}_q u_q \ \overline{u}_e \left[ F^{\mu e }_\gamma  \right] \gamma_\mu          P_L u_\mu\,,   \\
  i\mathcal{M}^q_Z      &=&   \frac{ i \alpha_W^2}{2 M^2_W} \overline{u}_q  \gamma^\mu  \left(\mathcal{I}^3_q P_L-\mathcal{Q}_q s^2_W \right) u_q \ \overline{u}_e \Big[ F^{\mu e }_Z   \Big] \gamma_\mu P_L u_\mu \, ,  \\
  i\mathcal{M}^q_{Box}  &=& \frac{i \alpha_W^2}{4M^2_W}  \overline{u}_q \gamma^\mu P_L u_q  \ \overline{u}_e \Big[F^{\mu e qq}_{Box} \Big] \gamma_\mu P_L u_\mu  \,. 
\end{eqnarray}
In terms of the above form factors, $\tilde{F}_u$ and $\tilde{F}_d$ as defined in  Eq.~(\ref{BRmueexact}) are given by 
\begin{eqnarray} \label{eqFtuApp}
  \tilde{F}_u^{\mu e} &=& \frac{2}{3} s^2_W \left( F^{\mu e }_\gamma-F^{\mu e }_Z\right)+ \frac{1}{4}  \left(  F^{\mu e }_Z + F^{\mu e uu}_{Box}\right)\, , \\ \label{eqFtdApp}
  \tilde{F}_d^{\mu e} &=&- \frac{1}{3} s^2_W \left( F^{\mu e }_\gamma-F^{\mu e }_Z\right)- \frac{1}{4}  \left(  F^{\mu e }_Z - F^{\mu e dd}_{Box}\right)\, . 
\end{eqnarray}
In the case of light nuclei, the contribution of the dipolar term can also be casted
in a four fermion interaction amplitude of the form
\begin{equation}
i\mathcal{M}^q_{\gamma-dipole} =  \frac{i\alpha_W \alpha }{2 M^2_W} \overline{u}_q  \gamma^\mu  \mathcal{Q}_q u_q \ \overline{u}_e \Big[-i\frac{\sigma_{\mu\nu}q^\nu}{q^2}\Big]m_\mu G_\gamma^{\mu e} P_L u_\mu  \, . \label{eqGgaApp}
\end{equation}
In this way Eq.~(\ref{BRmueapproxlight}) is obtained:
\begin{equation}
F_q^{\mu e}=Q_q s_W^2(F^{\mu e}_\gamma+G^{\mu e}_\gamma)+F^{\mu e}_Z\left(\frac{{\mathcal I}^3_q}{2}-Q_qs_W^2\right)+\frac{1}{4}F^{\mu eqq}_{box}\,.
\end{equation}

The loop functions entering the computation of the form factors in Eqs.~(\ref{eqFtuApp}-\ref{eqGgaApp}) are

\begin{eqnarray}
F_\gamma(x)&=& 	\frac{x(7x^2-x-12)}{12(1-x)^3} - \frac{x^2(x^2-10x+12)}{6(1-x)^4} \ln x	\, ,\\
G_\gamma(x)&=&    -\frac{x(2x^2+5x-1)}{4(1-x)^3} - \frac{3x^3}{2(1-x)^4} \ln x \, ,\label{Ggamma} \\
F_Z(x)&=& -\frac{5x}{2(1-x)}-\frac{5x^2}{2(1-x)^2}\ln x \, , \\
G_Z(x,y)&=& -\frac{1}{2(x-y)}\left[	\frac{x^2(1-y)}{1-x}\ln x - \frac{y^2(1-x)}{1-y}\ln y	\right]\, , \\
H_Z(x,y)&=&  \frac{\sqrt{xy}}{4(x-y)}\left[	\frac{x^2-4x}{1-x}\ln x - \frac{y^2-4y}{1-y}\ln y	\right] \, ,\\
F_{Box}(x,y)&=&\frac{1}{x-y}\Big\{\left(	4+\frac{xy}{4} 	\right) \left[\frac{1}{1-x}+\frac{x^2}{(1-x)^2} \ln x - \frac{1}{1-y}-\frac{y^2}{(1-y)^2} \ln y\right] \nonumber \\
 	 	&&-2 xy \left[\frac{1}{1-x}+\frac{x }{(1-x)^2} \ln x - \frac{1}{1-y}-\frac{y }{(1-y)^2} \ln y\right]\Big\} \, , \\
F_{XBox}(x,y)&=&\frac{-1}{x-y}\Big\{\left(	1+\frac{xy}{4} 	\right) \left[\frac{1}{1-x}+\frac{x^2}{(1-x)^2} \ln x - \frac{1}{1-y}-\frac{y^2}{(1-y)^2} \ln y\right] \nonumber \\
 	 	&&-2 xy \left[\frac{1}{1-x}+\frac{x }{(1-x)^2} \ln x - \frac{1}{1-y}-\frac{y }{(1-y)^2} \ln y\right]\Big\} \, ,
		\\ G_{Box}(x,y)&=&\frac{-\sqrt{xy}}{x-y}\Big\{\left(	4+xy 	\right) \left[\frac{1}{1-x}+\frac{x}{(1-x)^2} \ln x - \frac{1}{1-y}-\frac{y}{(1-y)^2} \ln y\right] \nonumber \\
 	 	&&-2 \left[\frac{1}{1-x}+\frac{x^2 }{(1-x)^2} \ln x - \frac{1}{1-y}-\frac{y^2 }{(1-y)^2} \ln y\right]\Big\} \, ,
 \end{eqnarray}

with the limiting values
\begin{align}
G_Z(0,x) & =  -\frac{x}{2(1-x)} \ln x  \, , & H_Z(0,x)  &= G_{Box}(0,x)=0 \, ,  \label{limitval1}\\
 F_{Box}(0,x)	 &  =    \frac{4}{ 1-x } + \frac{4x}{(1-x)^2}\ln x\, ,&
F_{XBox}(0,x)  &=   - \frac{1}{ 1-x } - \frac{x}{(1-x)^2}\ln x \, ,
\end{align}
\begin{align}
F_\gamma(x ) & \xrightarrow[x\ll 1]{}  -x \,, &  F_\gamma(x )  &  \xrightarrow[x\gg 1]{}    - \frac{7}{12}-\frac{1}{6} \ln x\, , \\
G_\gamma(x )  & \xrightarrow[x\ll 1]{}  \frac{x}{4}\, ,&  G_\gamma(x )  &  \xrightarrow[x\gg 1]{}    \frac{1}{2} \, , \label{Ggammaapprox}\\
F_Z(x) &  \xrightarrow[x\ll 1]{}    -\frac{5x}{2} \, ,  & F_Z(x )		  &  \xrightarrow[x\gg 1]{}   \frac{5}{2}-\frac{5}{2} \ln x\, , \\
G_Z(0,x ) & \xrightarrow[x\ll 1]{}  -\frac{1}{2} x \ln x
 \, , & G_Z(0,x ) & \xrightarrow[x\gg 1]{}  \frac{1}{2} \ln x  \, ,\\
 F_{Box}(0,x )& \xrightarrow[x\ll 1]{}  4\left(1+x\left(1+\ln x\right)\right)
  \, , & F_{Box}(0,x ) & \xrightarrow[x\gg 1]{}  0 \, , \\
 F_{XBox}(0,x )& \xrightarrow[x\ll 1]{}  -1-x\left(1+\ln x\right)
 \, , & F_{XBox}(0,x ) & \xrightarrow[x\gg 1]{}  0 \, .\label{limitval2}
\end{align}

In terms of these functions, the form factors read

\begin{eqnarray}
	 F^{\mu e }_\gamma &=& \sum_{i=1}^{3+k} U_{ei}U^*_{\mu i} F_\gamma(x_i)   = \sum_{i=1}^{k} U_{e N_i}U^*_{\mu N_i} F_\gamma(x_{N_i})\,,
	 \label{Fmue}\\
	 G^{\mu e }_\gamma &=& \sum_{i=1}^{3+k} U_{ei}U^*_{\mu i} G_\gamma(x_i)   = \sum_{i=1}^{k} U_{e N_i}U^*_{\mu N_i} G_\gamma(x_{N_i})\, ,  \label{Ggammamue} \\
	 F^{\mu e }_Z			 &=& \sum_{i,j=1}^{3+k} U_{ei}U^*_{\mu j} \left(\delta_{ij} F_Z(x_i) + C_{ij} G_Z(x_i,x_j) + C^*_{ij} H_Z(x_i,x_j)   \right) \\
	 &=& \sum_{i,j=1}^{k} U_{e N_i}U^*_{\mu N_j} \Big[\delta_{N_iN_j} \left(F_Z(x_{N_i}) +2 G_Z(0,x_{N_i})	\right) \nonumber \\
	 &&										 + C_{N_iN_j} \left(G_Z(x_{N_i},x_{N_j})- G_Z(0,x_{N_i})-G_Z(0,x_{N_j})	\right) + C^*_{N_iN_j}H_Z  (x_{N_i},x_{N_j}) \Big] \, , \\
 F^{\mu e uu}_{Box}&=&\sum_{i=1}^{3+k}\sum_{d_i=d,s,b} U_{ei}U^*_{\mu i} V_{u d_i} V^*_{u d_i} F_{Box}(x_i,x_{d_i}) \simeq \sum_{i=1}^{3+k} U_{ei}U^*_{\mu i} F_{Box}(x_i,0)  \\
	 &=&\sum_{i=1}^{k} U_{eN_i}U^*_{\mu N_i} \left[F_{Box}(x_{N_i},0)-F_{Box}(0,0) \right]\, ,\\
	 F^{\mu e dd}_{Box}&=&  \sum_{i=1}^{3+k}\sum_{u_i=u,c,t} U_{ei}U^*_{\mu i} V_{d  u_i } V^*_{d u_i} F_{XBox}(x_i,x_{u_i}) \simeq         \sum_{i=1}^{3+k} U_{ei}U^*_{\mu i}   F_{XBox}(x_i,0) \\
	 &=&\sum_{i=1}^{k} U_{eN_i}U^*_{\mu N_i} \left[F_{XBox}(x_{N_i},0)-F_{XBox}(0,0) \right]\, .
	 \label{Fmuedd}\\
	 	 F^{\mu eee}_{Box}&=&  \sum_{i,j=1}^{3+k} U_{ei}U^*_{\mu j}\left(U_{ei}U^*_{ej}G_{Box}(x_i,x_j)-2\,U^*_{ei}U_{ej}F_{XBox}(x_i,x_j)\right)  \\ \nonumber
	 &=&-2\sum_{i=1}^{k} U_{eN_i}U^*_{\mu N_i} \left[F_{XBox}(x_{N_i},0)-F_{XBox}(0,0) \right]\\ \nonumber
	 &&
	+\sum_{i,j=1}^{k}U_{eN_i}U^*_{\mu N_j}\Big\{ U_{eN_i}U^*_{eN_{j}}G_{Box}(x_{N_{i}},x_{N_j})-2\, U^*_{eN_i} U_{e N_j}\big[F_{XBox}(x_{N_i},x_{N_j})\\ &&- F_{XBox}(0,x_{N_j})
- F_{XBox}(x_{N_i},0)+F_{XBox}(0,0) \big]\Big\}
	 \label{Fmueee}
\end{eqnarray}
In the above, $x_{1,2,3}\equiv x_{\nu_{1,2,3}}\equiv m^2_{\nu_{1,2,3}}/M^2_W$,  $x_{4,...,3+k}\equiv x_{N_{1,...,k}}= m^2_{N_{1,...,k}}/M^2_W$, $x_q\equiv m^2_{q}/M^2_W$, $V$ is the quark CKM matrix and $U$ is the total $(3+k) \times (3+k)$ neutrino mixing matrix defined in Eq.~(\ref{Udef}). The second equality in Eqs.~(\ref{Fmue})-(\ref{Fmueee}) is obtained 
 using the unitarity identity $\sum_{i}U_{ei}U^*_{\mu i}=0$, the diagonalization relation $\sum_{i}U_{ei}\sqrt{x_{i}}\,U_{\mu i}=0$, the limiting
 values of the loop funtions in Eqs.~(\ref{limitval1})-(\ref{limitval2})  and the very good approximation for the present analysis in which light neutrino masses  are neglected  compared to the heavy neutrino masses  ($x_{N_{1,...,k}}\gg x_{\nu_{1,2,3}} $).

These form factors present the following behaviour for low and high masses
\begin{align}
	 F^{\mu e }_\gamma &  \xrightarrow[x\ll 1]{}   \sum_{i=1}^{k} U_{e N_i}U^*_{\mu N_i} \left[-x_{N_i}\right] ,&
	 	 F^{\mu e }_\gamma &  \xrightarrow[x\gg 1]{}   \sum_{i=1}^{k} U_{e N_i}U^*_{\mu N_i} \left[\frac{-7}{12}-\frac{1}{6} \ln x_{N_i}\right],  \\
	 G^{\mu e }_\gamma &  \xrightarrow[x\ll 1]{}  \sum_{i=1}^{k} U_{e N_i}U^*_{\mu N_i} \left[ \frac{x_{N_i}}{4} \right] , &
	G^{\mu e }_\gamma &  \xrightarrow[x\gg 1]{}  \sum_{i=1}^{k} U_{e N_i}U^*_{\mu N_i} \left[ \frac{1}{2} \right] , \\
	 F^{\mu e }_Z		
	 & \xrightarrow[x\ll 1]{} \sum_{i=1}^{k} U_{e N_i}U^*_{\mu N_i}x_{N_i}\left[  \frac{-5}{2}- \ln x_{N_i}\right] , &
	 	 F^{\mu e }_Z		
	 & \xrightarrow[x\gg 1]{} \sum_{i=1}^{k} U_{e N_i}U^*_{\mu N_i} \left[  \frac{5}{2}-\frac{3}{2} \ln x_{N_i}\right] , \\
	 F^{\mu e uu}_{Box}& \xrightarrow[x\ll 1]{}   \sum_{i=1}^{k} U_{eN_i}U^*_{\mu N_i}  4\,x_{N_i} \left[1+\ln x_{N_i}  \right]  ,&
	 	 F^{\mu e uu}_{Box}& \xrightarrow[x\gg 1]{}   \sum_{i=1}^{k} U_{eN_i}U^*_{\mu N_i} \left[ -4\right] , \\
	 F^{\mu e dd}_{Box}& \xrightarrow[x\ll 1]{}   \sum_{i=1}^{k} U_{eN_i}U^*_{\mu N_i}x_{N_i}\left[-1-\ln x_{N_i}  \right] ,&  
	 	 F^{\mu e dd}_{Box}& \xrightarrow[x\gg 1]{}    \sum_{i=1}^{k} U_{eN_i}U^*_{\mu N_i}\, ,\\
		 	 F^{\mu eee}_{Box}& \xrightarrow[x\ll 1]{}   \sum_{i=1}^{k} U_{eN_i}U^*_{\mu N_i}2\,x_{N_i}\left[1+\ln x_{N_i}  \right] ,&  
	 	 F^{\mu eee}_{Box}& \xrightarrow[x\gg 1]{}    \sum_{i=1}^{k} U_{eN_i}U^*_{\mu N_i}\,\left[ -2\right]\,  ,		 
\end{align}
where contributions involving 4 insertions of light-heavy mixing elements have been neglected; this is a  good approximation in 
view of the various experimental bounds which hold on the mixings for the low energy regime, whereas for very high right-handed masses this approximation
relies on the perturbativity of the yukawa couplings. Dropping these terms with four insertions,  the functions $\tilde{F}_q$ and $F_q$ can be written as
\begin{eqnarray}
\tilde{F}_u(x)&=&\ \ \frac{2}{3} s^2_W \Big[	F_\gamma(x) -F_Z(x) -2G_Z(0,x)	\Big] \nonumber \\
&&+ \frac{1}{4} 	\Big[F_Z(x) +2G_Z(0,x)+F_{Box}(x,0)-F_{Box}(0,0)\Big] \,  ,   \label{functionfutilde} \\ 
\tilde{F}_d(x)&=&-\frac{1}{3} s^2_W \Big[	F_\gamma(x) -F_Z(x) -2G_Z(0,x)	\Big] \nonumber \\
&&- \frac{1}{4} 	\Big[F_Z(x) +2G_Z(0,x)-F_{XBox}(x,0)+F_{XBox}(0,0)\Big] \,  ,     \label{functionfdtilde} \\ 
F_u(x)&=&\tilde{F}_u(x) +\frac{2}{3} s^2_W G_\gamma(x)\,  ,   \label{functionfu} \\
F_d(x)&=& \tilde{F}_d(x) -\frac{1}{3} s^2_W G_\gamma(x)\,  ,   \label{functionfd}
\end{eqnarray}
which are the expressions we actually use for all plots.

\end{document}